\documentclass[
reprint,
onecolumn,
nofootinbib,
amsmath,amssymb, aps
prd,
]{revtex4-2}
\pdfoutput=1

\usepackage[utf8]{inputenc}
\usepackage{graphicx}
\usepackage{amsmath}
\usepackage{amssymb}

\usepackage[usenames,dvipsnames]{xcolor}

\begin{document}

\title{Unified description of corpuscular and fuzzy bosonic dark matter}

\author{Nick P. Proukakis$^{1}$}
\email[Electronic mail: ]{nikolaos.proukakis@newcastle.ac.uk}
\author{Gerasimos Rigopoulos$^{1}$}
\email[Electronic mail: ]{gerasimos.rigopoulos@newcastle.ac.uk}
\author{Alex Soto$^{1}$}
\email[Electronic mail: ]{alex.soto@newcastle.ac.uk}
\affiliation{$^{1}$ School of Mathematics, Statistics and Physics, Newcastle University, Newcastle upon Tyne, NE1 7RU, UK}

\date{March 2023}

\begin{abstract}
{We derive from first principles equations for non-relativistic bosonic, self-interacting dark matter which can include both a condensed, low momentum ``fuzzy'' component and one with higher momenta that may be approximated as a collection of particles. The resulting coupled equations consist of a modified Gross-Pitaevskii equation describing the condensate and a kinetic equation describing the higher momentum modes, the ``particles'', along with the Poisson equation for the gravitational potential sourced by the density of both components. Our derivation utilizes the Schwinger-Keldysh path integral formalism and applies a semi-classical approximation which can also accommodate collisional terms amongst the particles and between the particles and the condensate to second order in the self-coupling strength. The equations can therefore describe both CDM and Fuzzy Dark Matter in a unified way, allowing for the coexistence of both phases and the inclusion of quartic self-interactions.}  
\end{abstract}

\maketitle

\section{Introduction}

Although the role of Dark Matter as one of the main architects of cosmic large scale structure is  well established, its physical nature is still practically unknown. The prevailing model of Cold Dark Matter (CDM) has been extremely successful in modelling the large-scale evolution and properties of the Universe \cite{Frenk-White_DM}, describing both the precise characteristics of the CMB temperature anisotropies as well as the late time large-scale mass distribution in both the linear and nonlinear regime. Formally, the CDM model consists of a collisionless Boltzmann kinetic equation for the CDM phase space density, coupled to the Poisson equation for the arising gravitational field. The N-Body solvers which underlie many of CDM's successful predictions in the non-linear regime are meant to sample that phase space density via the use of effective particles which trace it - see for example \cite{Bernardeau+, Angulo+Hahn}. 

Recently, an alternative idea for the microphysics of dark matter has been gaining traction among a growing number of cosmologists, and involves novel phenomenology on small scales - see \cite{2016PhR...643....1M, 2021ARA&A..59..247H, 2021A&ARv..29....7F} for recent reviews and references to the literature. Known as fuzzy dark matter (FDM) or $\psi$DM, it was put forward in \cite{Hu:2000ke} to address discrepancies between CDM and observations on sub-galactic length scales \cite{2015PNAS..11212249W, 2017ARA&A..55..343B, 2017Galax...5...17D} and over the past few years has started attracting increasing attention, especially after the pioneering simulations of \cite{2014NatPh..10..496S}. Indeed, an increasing number of ever more sophisticated simulations have been shedding light on the cosmological dynamics of fuzzy dark matter, see e.g. \cite{2016PhRvD..94d3513S, 2017MNRAS.471.4559M, Veltmaat:2018dfz, May_2021, Nori:2022afw,Mocz:2023adf}. The main idea is that the de Broglie wavelength of the constituent, bosonic particles can extend to galactic scales (indicatively $\lesssim$ 1 kpc) for an appropriately small particle mass: for DM halos hosting Milky Way sized galaxies, with corresponding typical velocities of $v \sim 250 \, km/s$, this mass would be around $10^{-22} eV/c^2$. This model exhibits wave-like properties at such cosmological scales and, unlike CDM, FDM is modeled by a Schr\"{o}dinger-type wave equation, coupled to the corresponding Poisson equation. Moreover, such particles may even exhibit a very weak self-interaction, modifying their primary underlying equation to a nonlinear Schr\"{o}dinger equation, making up the self-interacting fuzzy dark matter model \cite{Boehmer:2007um, Chavanis_2011,PhysRevD.84.043532,Rindler-Daller:2011afd, Guth:2014hsa, Chen:2021oot, Kirkpatrick:2021wwz, Hartman:2022cfh, Delgado:2022vnt, Mocz:2023adf}.

Dark matter made up of bosonic particles can fall anywhere between the two limiting cases of particle or wave-like behaviour. Given a velocity distribution, particles of different masses will exhibit wavy properties at different length scales, if at all, depending on their de Broglie wavelengths and number density. In particular, if their dimensionless phase space density   $(\rho / m) \lambda_{\rm dB}^3 \gtrsim O(1) $~\cite{2008bcdg.book.....P} their de Broglie wavelegths will exceed their typical inter-particle distance and the formation of a Bose-Einstein condensate is expected. For a DM halo hosting a Milky-Way sized galaxy this would occur for bosonic dark matter of $m \lesssim 30\, eV/c^2$ \cite{2021ARA&A..59..247H}. If DM is composed of bosons with masses greater than this value, it is expected that a corpuscular picture and a corresponding phase space density distribution suffices for the description of the dynamics. On the other hand, for lower masses, we are in the realm of wave dark matter, usually described by the corresponding wave equation. More precisely, condensation in a homogeneous, non-interacting 3D gas in thermal equilibrium occurs when
\begin{equation}\label{cond_crit-1}
{\cal D}=\frac{\rho}{m} \lambda_\mathrm{dB}^3
=\frac{h^3}{m^4} \frac{\langle\rho\rangle}{\langle|{v}|\rangle^3} \gtrsim 
\zeta(3/2) \approx 2.612\,.
\end{equation}
which simply states the necessity to have more than approximately one particle per de Broglie volume. As such, this is a general criterion that can also be used in a dark matter halo consisting of scalar, bosonic particles of mass $m$ in which case both $\rho$ and $v$ are spatially dependent and there is no thermal equilibrium. 
In the latter case, condensation will thus occur when   
\begin{equation}
{\cal D}(r)
=\left(\frac{h^3\rho_\mathrm{ref}}{m^4v_\mathrm{ref}^3}\right)\frac{\langle\rho^\prime(r)\rangle}{\langle|{v}^\prime(r)|\rangle^3}={\cal D}_\mathrm{ref}\,\frac{\langle\rho^\prime(r)\rangle}{\langle|{v}^\prime(r)|\rangle^3} \gtrsim O(1) \;.
\end{equation}
Here primes refer to dimensionless quantities which can be accessed via numerical simulation, and  
\begin{equation}
{\cal D}_\mathrm{ref}=\displaystyle\frac{h^3\rho_\mathrm{ref}}{m^4v_\mathrm{ref}^3} \;.
\end{equation}
gives a constant reference phase space density which can be expressed as:
\begin{equation}
{\cal D}_\mathrm{ref}\approx \displaystyle1.25\times10^{90}\left(\frac{10^{-22}\textrm{eV}/c^2}{m}\right)^4\left(\frac{\rho_\mathrm{ref}}{10^3M_\odot\mathrm{kpc}^{-3}}\right)\left(\frac{250\mathrm{km/s}}{v_\mathrm{ref}}\right)^3
\end{equation} 
The value of $250$ $km/s$ is chosen as the typical velocity of a particle in a DM halo hosting a Milky-way-like galaxy. We see that the strongest influence on ${\cal D}_\mathrm{ref}$ comes from the boson mass value and that to bring ${\cal D}_\mathrm{ref} \sim \mathcal{O}(1)$ we need $m\sim \mathcal{O}(1) \, eV/c^2$. However, the radial profile of ${\cal D}(r)$ can also have an effect as the density is expected to vary by some orders of magnitude from the centre to the edge of a halo. Hence, for masses relatively close to the $eV$ regime, one may expect configurations where partially condensed regions can coexist with the uncondensed phase. Furthermore, even in the lower mass, fuzzy dark matter limit where ${\cal D}$ is very large and the use of a wave equation may be fully justified, there are important {\it qualitative differences} between the field configuration in the central soliton and the outside halo \cite{2022arXiv221102565L}. While the solitonic core is completely coherent and has the characteristics of a true condensate, the surrounding halo is on average incoherent and rather turbulent. It may thus be possible that the highly dynamical halo can be coarsely described as a collection of ``particles", even for the high occupation numbers corresponding to the lower bosonic masses which validate a classical field description. We therefore conclude that there must be regimes where either: a) a mixture of condensed and non-condensed phases of the same underlying DM bosonic field physically coexist or b) a ``condensate - particles'' division may offer a convenient approximate way to describe the full, classical field dynamics.  

It is important to remember that whether bosonic dark matter behaves as a wave (BEC) or a collection of particles is not a fundamental property of the underlying physical model, which is a bosonic field of fundamental mass $m$, but of the particular configuration and the available energy, more like different phases of the same underlying quantum field. In this work we obtain equations that can describe both phases, either in isolation or coexistence, from first principles. We start from the underlying, fundamental non-equilibrium Schwinger-Keldysh path integral for a non-relativistic, bosonic field of mass $m$ coupled to gravity and derive a collisional kinetic equation for the particle-like component along with a nonlinear Schr\"{o}dinger equation for the condensate. These are coupled via the field's quartic self-interaction, along with the Poisson equation sourced by the density of the two phases which determines their common gravitational field.

\begin{figure}[t]
    \centering
    \includegraphics[width=1\linewidth,]{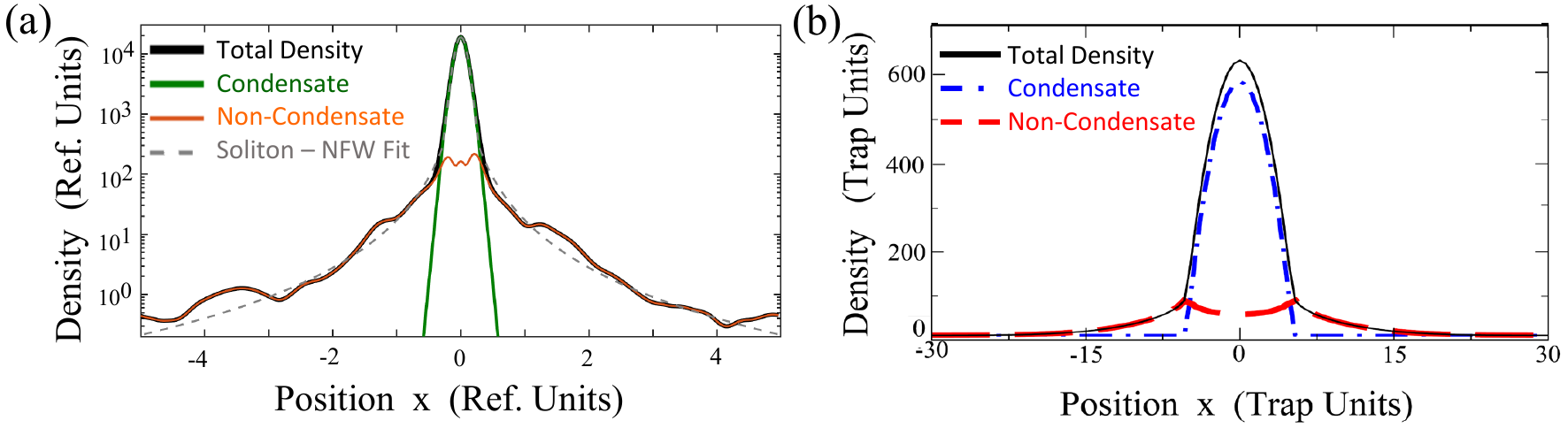}
    \caption{Similarity of the respective densities in the context of (a) gravitationally-attractive non-self-interacting FDM and (b) weakly-interacting harmonically-trapped cold atomic gases exhibiting partial BEC:
    in both cases, numerical simulations identify a broad total density distribution (black lines) featuring a dominant coherent central region representing the BEC, surrounded by an incoherent field configuration of particles representing the non-BEC.
    The FDM density distribution features the characteristic soliton expected in the centre of the gravitational potential of a halo,
    with averaged density outside the solitonic core following the expected NFW profile, also predicted by corpuscular N-body simulations of CDM. In this case, the BEC is identified as the mode corresponding to the largest eigenvalue of the one-body density matrix (the so-called Penrose-Onsager mode), with the rest of the particles labelling the incoherent field: depicted results correspond to the data of Liu {\em et al.}~\cite{2022arXiv221102565L} based on numerical solution of the Schr\"{o}dinger-Poisson equations, and are plotted in terms of appropriate length and density units (see above work).
    In the atomic case the surrounding cloud is a collection of harmonically-trapped thermal particles following the Maxwell-Boltzmann distribution: data based on the self-consistent Hartree-Fock theory (see, e.g.~\cite{proukakis2008finite}). 
    Remarkably, in both (a) virialized FDM and (b) equilibrated cold-atom cases there is a centrally-peaked coherent feature, with the incoherent component still present in that region: the small localised central dip in the incoherent atomic density is due to (weak) repulsive interactions between atoms (and between the coherent and incoherent parts), with the incoherent component having a Gaussian distribution (also) at the centre in the absence of such interactions. (The presented FDM results do not account for any interactions.) Note that the FDM density axis is plotted on a logarithmic scale, so the gravitational attraction leads to an extremely high concentration of central density and corresponding degree of coherence of the total system density within the soliton (due to the high BEC to total density fraction), whereas the extent of the centrally-located incoherent feature (and thus the central condensate density fraction) in the atomic case depends on the ratio of the system temperature, to the corresponding critical temperature for BEC. The analogy shown here between a gravitationally-bound cored-halo and a harmonically-trapped cold atomic gas was first pointed out in Ref.~\cite{2022arXiv221102565L}. 
    }
    \label{fig:1}
\end{figure}

Our approach draws inspiration from other fields of physics where bosonic condensates and incoherent particles can coexist: 
The best known example of this arises in the case of superfluid liquid helium, where the superfluid component coexists with the (usual) normal fluid component in thermodynamic equilibrium at a temperature below some critical value (arising at the so-called $\Lambda$ critical point), in the context of the established two-fluid model~\cite{tisza-2-fluid,landau-2-fluid,griffin_nikuni_zaremba_2009}.
Nonetheless, the nature and strength of particle interactions in such systems leads to a distinction between the superfluid and the condensate fraction, with liquid helium exhibiting less than 10\% condensate fraction, even in a pure ($T=0$) superfluid~\cite{Penrose-Onsager}.
Moreover, the typical spatial homogeneity of the box renders this a rather partial analogy to the cosmological setting of inhomogeneous density, with the significance of inhomogeneous confinement evident in the discussion below. During the past 3 decades, other more weakly-interacting systems have come to the forefront, whose weakness of interactions facilitates a more direct dilute Bose gas 
description.
The simplest such configuration arises in optically/magnetically trapped atomic gases, thermalized below some characteristic temperature~\cite{Cornell-RevModPhys.74.875,Ketterle-RevModPhys.74.1131,Dalfovo1999}, but similar condensation features can also be found in the two-dimensional context of photons weakly interacting (indirectly) with each other within a dye medium~\cite{Klaers2010}, or driven-dissipative exciton-polariton condensates in semiconductor microcavities above some pumping threshold~\cite{Kasprzak2006}.

Here we focus for concreteness on the case of ultracold atomic condensates which (unlike photon, or exciton-polariton condensates) also exist in three dimensions:
in such systems, the typical existence of an external magnetic/optical confining potential leads to centrally-peaked inhomogeneous density profiles, which thus have some qualitative analogy with the (gravitationally-bound) density-peaked central regions of cosmological halos.
Such analogy is shown in Fig.~\ref{fig:1} between the density profiles of a cosmological (cored, fuzzy) dark matter halo (left) and the corresponding atomic gas density distribution (right) based on realistic numerical simulations, with the depicted densities integrated over the other two (transverse) directions in both cases. 
In the well-studied atomic gas case (Fig.~\ref{fig:1}(b)), one routinely identifies a coherent  component (condensate) residing in the centre of the trap (dash-dotted blue line), 
surrounded by a cloud of incoherent (non-condensate) particles (dashed red line)~\cite{proukakis2008finite}; moreover, the two components co-exist in the trap centre (with the small dip in the thermal cloud density in such overlap region arising from repulsive atomic interactions). 
Such a picture is fully established and has been directly confirmed in a range of experiments, with such `bimodal' fits a standard tool in the ultracold atomic community.
Remarkably, this situation is somewhat analogous to the rich spatial profile exhibited by gravitationally-bound virialized solitonic cores, known to form in fuzzy dark matter, and their surrounding incoherent halo.
In the latter context, and building on an extensive body of earlier work analyzing such density profiles~\cite{2014NatPh..10..496S,2016PhRvD..94d3513S,2017MNRAS.471.4559M,Veltmaat:2018dfz}, Liu et al \cite{2022arXiv221102565L} significantly augmented such picture by explicitly demonstrating the continuous decrease of coherence from the inner coherent solitonic core (the `condensate', solid green line in Fig.~\ref{fig:1}(a)) which exhibits a very high phase-space density ($\gg O(1)$),  through to the outer `incoherent' regions of much lower-density which follow the NFW density profile: 
in fact the entire FDM profile can be excellently fitted both by the combination of a central soliton core and a larger-distance NFW profile (dashed grey line)~\cite{2014NatPh..10..496S,2017MNRAS.471.4559M,2018MNRAS.478.2686C,2021PhRvD.103j3019C,2022arXiv221102565L}, and by the sum of a condensate (solid green line) and a non-condensate (thin orange line) component~\cite{2022arXiv221102565L}.
Even in such a cored-halo case, the coherent and incoherent components were also numerically found to co-exist in the central region, although the condensate core is here significantly more dominant than the incoherent component in the central region when compared to the atomic gas case (note the logarithmic scale in Fig.~\ref{fig:1}(a)) -- a direct consequence of the extremely high condensate fraction in the central region due to the strong gravitational attraction.
Moreover, the outer halo regions were found to support a dynamically-evolving turbulent, quantum vortex tangle delineating intermediate-scale `granular' regions of relatively suppressed density fluctuations~\cite{2022arXiv221102565L}, with such features washing away after extensive time averaging to yield the averaged profile shown in Fig.~\ref{fig:1}(a). 

In the context of the above mentioned atomic systems -- which have been found to exhibit
both weak (wave) turbulence and strong turbulence -- one typically uses a description based on a nonlinear Schr\"{o}dinger equation, known as the Gross-Pitaevskii equation, appropriately generalized to include the coupling of the coherent and incoherent parts of the system through a variety of approaches.
Various such theoretical models used in ultracold quantum gases have been summarized in Refs.~\cite{proukakis2008finite,Proukakis13Quantum,berloff_brachet_14} and broadly follow very distinct, yet notionally rather related, approaches based on effective field theory~\cite{Stoof-PhysRevLett.78.768,stoof1999coherent,kamenev_2011}, kinetic theories~\cite{Kirkpatrick-Dorfman-83-PhysRevA.28.2576,Kirkpatrick-Dorfman-1-JLTP-85,Kirkpatrick-Dorfman-2-JLTP-85,zaremba1999dynamics,zaremba-stoof-growth-PhysRevA.62.063609,griffin_nikuni_zaremba_2009} or quantum-optical  approaches~\cite{Blakie-AdvPhys-2008}. The underlying features lie in the description of either a highly-occupied classical multi-mode system encompassing condensate plus low-lying excitations~\cite{Blakie-AdvPhys-2008}, or a purely condensed component~\cite{zaremba1999dynamics},  respectively coupled to the higher-lying modes in the system, which, in their most complete model~\cite{stoof1999coherent,zaremba-stoof-growth-PhysRevA.62.063609,duine-stoof-PhysRevA.65.013603,griffin_nikuni_zaremba_2009} are themselves coupled to appropriate kinetic equations (although numerical implementations of effective field theories treat this as an effective bath~\cite{stoof_dynamics_2001,duine-stoof-PhysRevA.65.013603,proukakis2008finite,Blakie-AdvPhys-2008,Proukakis13Quantum,liu_dynamical_2018}). Such models are also closely related to a description in the context of a single classical field equation~\cite{kagan-qc-growth,kagan-greenbook}, typically starting from an initial non-equilibrium configuration~\cite{Berloff2002a,Blakie-AdvPhys-2008}.
In fact, the latter equation also reduces, in appropriate limits and semi-classical assumptions, to a semi-classical Boltzmann-type equation~\cite{kagan-greenbook,connaughton-pomeau,Proukakis13Quantum} - see also related work on condensation of bosons within (restricted) applicable regimes~\cite{semikoz-tkachev-PhysRevLett.74.3093}.
Such latter analogy between underlying theoretical approaches directly connects the modelling of quantum coherent systems to wave turbulence phenomena.
The existence and interconnectedness of such seemingly very diverse approaches -- whose power in describing ultracold quantum gases is directly verified by their successful modelling of a broad range of experimental observations -- demonstrates an inherent flexibility of describing a physical system by different means, which nonetheless lead to the same results in relevant regimes of applicability~\cite{Proukakis13Quantum}.

The aim of this work is to provide a corresponding overarching theoretical framework, underpinning the study of different dark matter models, which is motivated by considerations like the above, is consistent both with a kinetic equation and with wave turbulence, and directly reduces to the established CDM and $\psi$DM models in appropriate limits. Equipped with the understanding of such diverse models, we use a non-equilibrium field-theoretic approach (the Schwinger-Keldysh or in-in formalism) to derive a general framework of coherent and incoherent modes in a cosmological context\footnote{The derivation is similar in spirit to \cite{Friedrich:2019zic} which however did not involve self-interactions or the possible existence of a condensate.}. We will arrive at a general cosmological model that encompasses both the collisionless Boltzmann equation of CDM (in the limit of no condensate particles) and the $\psi$DM model (when the incoherent density is ignored). Our formalism therefore provides a more general framework for studying different dark matter models, with the different established frameworks described by a subset of our equations. 
Future work will address simulations of our equations in an attempt to test subtle aspects in the applicability regimes of each model, whether such a combined model may be required for some astrophysical observations and, if so, what constraints may be placed on the combination of mass and interaction strength. Furthermore, in this work we focus on the leading, semi-classical dynamics of the condensate $\Phi$, and the gravitational potential $V$, described in the action by terms of linear order in the ``Keldysh fields'' $\Phi^q$ and $V^q$. A forthcoming publication will treat fluctuations and noise, coming from terms of higher order in these fields.

It is of course expected that certain spatial regions of the scalar dark matter configuration may be adequately described by either the coherent or the incoherent modes alone. A likely example could be the central regions of the solitonic cores that form in halos, where the very high condensate fraction would indicate that the role of the incoherent part is likely to be small there (and potentially negligible). Moreover, the incoherent modes may be an adequate description of the outer parts of the halos. Our extended model can offer a unified, self-consistent description, and with minimal further assumptions, of a spatially-varying physical configuration, smoothly interpolating as a function of position between the FDM (solitonic core) and CDM descriptions, without the explicit need for a potentially {\em ad hoc} spatial separation between the two regions.

The paper is structured as follows: Section 2 describes the Schwinger-Keldysh incarnation of the non-relativistic action for a boson of mass $m$, interacting via gravity and a quartic self-coupling, and the split into a ``slow'' (condensate) and a ``fast'' (particle) component. We derive an effective action for the condensate via integrating out the particles. Equations for the fast component are also described: these are the Schwinger-Dyson equations for its 2-point function, or the Kadanoff-Baym equations as they are known in the Schwinger-Keldysh, non-equilibrium context. Section 3 then obtains the fundamental dynamical equations for these two components, namely a Gross-Pitaevskii-type equation for the condensate and a kinetic equation for the particles, derived from the two point function by utilizing the leading order of a Wigner transform. In both equations we include collisional terms up to second order in the self-coupling. Importantly, we also discuss the various limits of these equations and how they encompass existing/established dark matter models. We close with a discussion of our results and some thoughts of their relation to the wider landscape of dark matter modeling in section 4. Some technical details, including the computation of the collisional terms at order in the self coupling and up to two loops, are relegated to two appendices.

\section{Model and action for the boson}

Our starting point is the non-relativistic action for a boson of mass $m$ in
an expanding universe given by
\begin{equation}
\label{eqini}
  S = \int d t d^3 x \left( i \phi^{\ast} \partial_t \phi + \frac{1}{2
  m a^2} \phi^{\ast} \nabla^2 \phi - \frac{g}{2 a^3} | \phi |^4 - m V | \phi
  |^2 - \frac{a}{8 \pi G} (\nabla V)^2 \right)
\end{equation}
where we have set $\hbar=1$ for simplicity.  Here, $\phi$ is the wavefunction of the boson, $a=a(t)$ is the scale factor
describing the expansion of the universe, $g$ is the coupling of the boson
self-interaction and $V$ corresponds to the gravitational potential. This action can be obtained from the relativistic action of a minimally coupled scalar field with gravity under the assumptions of weak gravitational field, slow velocities and a posterior redefinition of the scalar field as $\phi\to a^{-3/2}\phi$ which re-absorbs the derivatives of the scale factor. 

We will work in the Schwinger-Keldysh or ``in-in'' formalism \cite{1986PhRvD..33..444J, 1988PhRvD..37.2878C, Stoof-PhysRevLett.78.768, stoof1999coherent,kamenev_2011, RammerBook}  which allows the computation of in-in expectation values and results in real equations of motion for them, unlike the more commonly used ``in-out'' formalism which is more appropriate for the computation of scattering amplitudes. The Schwinger-Keldysh formalism involves an integration forwards and backwards in time which in practice can be thought of as a single integration but with double the degrees of freedom in the fields, usually denoted as $\phi_+$ and $\phi_-$ on the forward and backward time contour, along with the extra condition that $\phi_+(t_f)=\phi_-(t_f)$ at some final time $t_f$ in the far future. Furthermore, performing a Keldysh rotation 
\begin{equation}
    \phi^{cl} = \frac{\phi_+ + \phi_-}{\sqrt{2}}\,, \quad\phi^{q} = \frac{\phi_+ - \phi_-}{\sqrt{2}}
\end{equation}
and 
\begin{equation}
    V^{cl} = \frac{V_+ + V_-}{2}\,, \quad V^{q} = \frac{V_+ - V_-}{2}
\end{equation}
the Keldysh action for the theory (\ref{eqini}) can then be written as
\begin{eqnarray}
\label{actionsk}
  S &=& S[\phi_+,V_+] - S[\phi_-,V_-]\\
    &=& \int d t d^3 x \, \phi^{q \ast} \left[ i \partial_t
  + \frac{\nabla^2}{2 m a^2} \right] \phi^{cl} + \int dt d^3 x  \, \phi^{cl \ast} \left[ i \partial_t + \frac{\nabla^2}{2 m a^2} \right] \phi^q \nonumber\\
& & - \frac{g}{2a^3} \int d t d^3 x (\phi^{q \ast} (|
  \phi^{cl} |^2 + | \phi^q |^2) \phi^{cl} + \phi^{cl
  \ast} (| \phi^{cl} |^2 + | \phi^q |^2) \phi^q) \nonumber\\
& & + \frac{1}{4 \pi G} \int d t d^3 x\,  aV^q \nabla^2
  V^{cl} + \frac{1}{4 \pi G} \int d t d^3 x \,a 
  V^{cl} \nabla^2 V^q \nonumber\\
& & - m \int d t d^3 x \, (\phi^{cl \ast} V^q \phi^{cl}
  + \phi^{cl \ast} V^{cl} \phi^q + \phi^{q \ast} V^{cl}
  \phi^{cl} + \phi^{q \ast} V^q \phi^q)\label{KeldyshAction}
\end{eqnarray}
where the minus sign in (\ref{actionsk}) comes from the backwards temporal integral and in (\ref{KeldyshAction}) the action is expressed in terms of the Keldysh ``classical'' and ``quantum'' fields. All quantities of interest and the main equations of our model can then be obtained from the generating functional
\begin{equation}\label{eq:PartitionFn}
Z = \int \mathcal{D} [\phi_+ \phi_- V_+ V_-] \, e^{i \left(S_+-S_-\right)} =\int \mathcal{D} [\phi^{cl} \phi^q V^{cl} V^q] \, e^{i S}
\end{equation}
where $S$ in the exponent is given by the Keldysh form of the action, given in (\ref{KeldyshAction}).  

To obtain the dynamical equations for our ``condensate + particles" system we will split the fields in a ``slow'' part,
which we will associate with the condensate and a ``fast'' one, which we will associate with the corpuscular component. In this paper we will be interested in a
semiclassical description of the condensate and the gravitational potential
$V$. For this reason, it is enough to keep terms up to first order in the quantum component of the slow part of the bosonic field ($\Phi^q$) and the gravitational potential ($V^q$), neglecting any higher orders in these quantities. However, we will keep higher orders in the fast part of the boson as we will be integrating out fluctuations of the fast fields. In addition, we will consider the Popov approximation to simplify some
computations - see eg \cite{kamenev_2011}. 

\subsection{Splitting in slow and fast parts}
We split the classical and quantum fields in equation \eqref{actionsk} in terms of a slow part (the condensate) and a fast one (the particles): $\phi^{cl} = \Phi_0 +
\varphi$ and $\phi^q = \Phi^q + \varphi^q$. Explicitly, we have for the Keldysh action
\begin{eqnarray}
\label{action1}
  S &=& \int d t d^3x \left(\begin{array}{cc}
    \Phi_0^{\ast} & \Phi^{q \ast}
  \end{array}\right) \left(\begin{array}{cc}
    0 & i \partial_t - H_c\\
    i \partial_t - H_c & 0
  \end{array}\right) \left(\begin{array}{c}
    \Phi_0\\
    \Phi^q
  \end{array}\right)\nonumber\\ 
& & + \frac{1}{2} \int d t  d^3x \left(\begin{array}{cc}
    \xi^{\dag} & \xi^{q \dag}
  \end{array}\right) \left(\begin{array}{cc}
    0 & M\\
    M & 0
  \end{array}\right) \left(\begin{array}{c}
    \xi\\
    \xi^q
  \end{array}\right)\nonumber\\
& & + \frac{a}{4 \pi G} \int d t  d^3x \left(\begin{array}{cc}
    V^{cl} & V^q
  \end{array}\right) \left(\begin{array}{cc}
    0 & \nabla^2\\
    \nabla^2 & 0
  \end{array}\right) \left(\begin{array}{c}
    V^{cl}\\
    V^q
  \end{array}\right) \nonumber\\
& & - m \int d t  d^3x \,V^q (| \Phi_0 |^2 + \varphi^{\ast} \varphi
  + \varphi^{q \ast} \varphi^q) \nonumber\\
& & - \frac{g}{2 a^3} \int d t  d^3x \bigg( 2 \Phi^{q \ast} \Phi_0
  \varphi^{\ast} \varphi - 2 \Phi^{q \ast} \Phi_0 \langle \varphi^{\ast} \varphi \rangle + 2 \Phi^{q \ast} \Phi_0 \varphi^{q \ast} \varphi^q +
  \Phi^{q \ast} \Phi^{\ast}_0 \varphi \varphi + \Phi^{q \ast} \Phi^{\ast}_0
  \varphi^q \varphi^q \nonumber\\
& & + 2 \Phi^{\ast}_0 \Phi^q \varphi^{\ast} \varphi - 2 \Phi^{\ast}_0 \Phi^q \langle \varphi^{\ast} \varphi \rangle + 2
  \Phi^{\ast}_0 \Phi^q \varphi^{q \ast} \varphi^q + \Phi_0 \Phi^q
  \varphi^{\ast} \varphi^{\ast} + \Phi_0 \Phi^q \varphi^{q \ast} \varphi^{q
  \ast} \bigg) \nonumber\\
& & - \frac{g}{2 a^3} \int d t  d^3x \bigg( \Phi^{q \ast}
  \varphi^{\ast} \varphi \varphi + 2 \Phi^{q \ast} \varphi^{q \ast} \varphi^q
  \varphi + \Phi^{q \ast} \varphi^{\ast} \varphi^q \varphi^q + \Phi^q
  \varphi^{\ast} \varphi^{\ast} \varphi \nonumber\\
& & + 2 \Phi^q \varphi^{\ast} \varphi^{q
  \ast} \varphi^q + \Phi^q \varphi^{q \ast} \varphi^{q \ast} \varphi + 2
  \Phi_0 \varphi^{q \ast} \varphi^{\ast} \varphi + \Phi_0 \varphi^{\ast}
  \varphi^{\ast} \varphi^q + \Phi_0 \varphi^{q \ast} \varphi^{q \ast}
  \varphi^q \nonumber\\
& & + 2 \Phi^{\ast}_0 \varphi^{\ast} \varphi \varphi^q + \Phi^{\ast}_0
  \varphi^{q \ast} \varphi \varphi + \Phi^{\ast}_0 \varphi^{q \ast} \varphi^q
  \varphi^q \bigg)\nonumber\\
& & - \frac{g}{2 a^3} \int d t  d^3x \bigg(\varphi^{q \ast}
  \varphi^{\ast} \varphi \varphi + \varphi^{\ast} \varphi^{q \ast} \varphi^q
  \varphi^q + \varphi^{\ast} \varphi^{\ast} \varphi \varphi^q + \varphi^{q
  \ast} \varphi^{q \ast} \varphi^q \varphi \nonumber\\
  & & - 2 \langle \varphi^{\ast} \varphi \rangle \varphi^{\ast} \varphi^q - 2
\langle \varphi^{\ast} \varphi \rangle \varphi^{q \ast} \varphi\bigg)
\end{eqnarray}
where we have defined the condensate Hamiltonian
\begin{equation}
H_c= -\frac{1}{2 m a^2} \nabla^2 + m V^{cl} + \frac{g}{2 a^3} |\Phi_0 |^2 + \frac{g}{a^3} \langle \varphi^{\ast} \varphi \rangle
\end{equation}
and
\begin{equation}
\xi = \left(\begin{array}{c}
  \varphi\\
  \varphi^{\ast}
\end{array}\right), \qquad \xi^q = \left(\begin{array}{c}
  \varphi^q\\
  \varphi^{q \ast}
\end{array}\right), \qquad M = \left(\begin{array}{cc}
  i \partial_t - H_{qp} & - \frac{g}{2 a^3} \Phi_0 \Phi_0\\
  - \frac{g}{2 a^3} \Phi^{\ast}_0 \Phi^{\ast}_0 & - i \partial_t - H_{qp}
\end{array}\right)
\end{equation}
with the ``quasiparticle'' Hamiltonian\footnote{Here we adopt the  language found e.g.~in Kamenev's book~\cite{kamenev_2011}, noting for completeness that, since the implied dispersion relation is quadratic in $k$, this is technically a dressed {\em single-particle} Hamiltonian, reminiscent of the Hartree-Fock limit routinely used in ultracold quantum gases~\cite{proukakis2008finite,griffin_nikuni_zaremba_2009}.} given by
\begin{equation}
H_{qp} = - \frac{1}{2 m a^2} \nabla^2 + m V^{cl} + \frac{g}{a^3} | \Phi_0 |^2 + \frac{g}{a^3} \langle \varphi^{\ast} \varphi \rangle \;.
\end{equation}
As we mentioned above, we are only keeping to first order in the quantum fields $\Phi^q$ and $V^q$ since we are interested in a semiclassical description of the condensate and the gravitational potential. The next order in the tree-level action involves cubic terms in these q-fields. In contrast, all terms are kept for the fast, quasi-particle fields $\varphi$. Furthermore, we have placed the slow terms $V^{cl}$ and $| \Phi_0 |^2$ inside the Hamiltonians as parts of an effective potential. As we have introduced an order $g$ term as part of a mean-field approximation, we must consider any other term of order $g$ coming from corrections due to fluctuations of the particle field $\varphi$. For this reason, we have added the term $\frac{g}{a^3} \langle \varphi^{\ast} \varphi \rangle$ in the condensate and particle Hamiltonians, also subtracting the same quantity from the interaction terms. This arrangement will be useful later.

In the above grouping of the interaction terms it is important to highlight that terms with a single $\varphi$ field have been omitted as they would not contribute to viable vertices given our assumption that $\Phi$ and $\varphi$ correspond to ``slow'' and ``fast'' fields. Firstly, terms with one
condensate and one $\varphi$ as well as terms with three condensates and one $\varphi$
cannot be present since such combinations would violate conservation of
energy-momentum. 
Secondly, terms with one condensate, one $V$ and a number of $\varphi$'s don't appear since
the gravitational potential doesn't change the fastness or slowness of a particle.

In order to simplify the computations we can make the (rather common) approximation of dropping the off-diagonal terms in the matrix $M$ (usually motivated on the grounds that such terms are smaller, and/or evolve faster than other slow variables of our choice). 
This is equivalent to the so-called Popov approximation, which posits that $\langle \varphi
\varphi \rangle$, $\langle \varphi^{\ast} \varphi^{\ast} \rangle$, $\langle
\varphi^q \varphi^q \rangle$ and $\langle \varphi^{q \ast} \varphi^{q \ast}\rangle$ can be neglected. 
In principle, such terms could also be formally incorporated via diagonalisation of the matrix $M$. For more details we refer the reader to appendix \ref{appA}. With this, we can write the action as
\begin{equation}
S=S_0+S_I
\end{equation}
where we have further split the action in two pieces defined as
\begin{eqnarray}
\label{effaction0}
  S_0 &=& \int d t d^3 \mathbf{r} \left(\begin{array}{cc}
    \Phi_0^{\ast} & \Phi^{q \ast}
  \end{array}\right) \left(\begin{array}{cc}
    -m V^q & i \partial_t - H_c\\
    i \partial_t - H_c & 0
  \end{array}\right) \left(\begin{array}{c}
    \Phi_0\\
    \Phi^q
  \end{array}\right) \nonumber\\
& & + \int d t d^3 \mathbf{r} \left(\begin{array}{cc}
    \varphi^{\ast} & \varphi^{q \ast}
  \end{array}\right) \left(\begin{array}{cc}
    -m V^q & i \partial_t - H_{qp}\\
    i \partial_t - H_{qp} & -m V^q
  \end{array}\right) \left(\begin{array}{c}
    \varphi\\
    \varphi^q
  \end{array}\right) \nonumber\\
& & + \frac{1}{4 \pi G} \int d t d^3 \mathbf{r} \left(\begin{array}{cc}
    V^{cl} & V^q
  \end{array}\right) \left(\begin{array}{cc}
    0 & a\nabla^2\\
    a\nabla^2 & 0
  \end{array}\right) \left(\begin{array}{c}
    V^{cl}\\
    V^q
  \end{array}\right)
\end{eqnarray}
and
\begin{eqnarray}
\label{effaction0b}
  S_I &=& - \frac{g}{2} \int \frac{ d t d^3\mathbf{r}}{a^3}  \bigg( 2 \Phi^{q \ast} \Phi_0
  \varphi^{\ast} \varphi - 2 \Phi^{q \ast} \Phi_0 \langle \varphi^{\ast} \varphi \rangle + 2 \Phi^{q \ast} \Phi_0 \varphi^{q \ast} \varphi^q + \Phi^{q
  \ast} \Phi^{\ast}_0 \varphi \varphi + \Phi^{q \ast} \Phi^{\ast}_0 \varphi^q \varphi^q \nonumber\\
& & + 2 \Phi^{\ast}_0 \Phi^q \varphi^{\ast} \varphi - 2 \Phi^{\ast}_0 \Phi^q \langle \varphi^{\ast} \varphi \rangle + 2 \Phi^{\ast}_0 \Phi^q \varphi^{q
  \ast} \varphi^q + \Phi_0 \Phi^q \varphi^{\ast} \varphi^{\ast} + \Phi_0 \Phi^q \varphi^{q
  \ast} \varphi^{q \ast} \bigg) \nonumber\\
& & - \frac{g}{2} \int \frac{ d t d^3\mathbf{r}}{a^3}  \bigg( \Phi^{q \ast}
  \varphi^{\ast} \varphi \varphi + 2 \Phi^{q \ast} \varphi^{q \ast} \varphi^q \varphi + \Phi^{q
  \ast} \varphi^{\ast} \varphi^q \varphi^q + \Phi^q \varphi^{\ast} \varphi^{\ast} \varphi \nonumber\\
& & + 2 \Phi^q \varphi^{\ast} \varphi^{q \ast} \varphi^q + \Phi^q \varphi^{q \ast} \varphi^{q \ast}
  \varphi + 2 \Phi_0 \varphi^{q \ast} \varphi^{\ast} \varphi + \Phi_0 \varphi^{\ast}
  \varphi^{\ast} \varphi^q + \Phi_0 \varphi^{q \ast} \varphi^{q \ast} \varphi^q \nonumber\\
& & + 2 \Phi^{\ast}_0 \varphi^{\ast} \varphi \varphi^q + \Phi^{\ast}_0 \varphi^{q \ast} \varphi
  \varphi + \Phi^{\ast}_0 \varphi^{q \ast} \varphi^q \varphi^q \bigg) \nonumber\\
& & - \frac{g}{2}\int \frac{ d t d^3\mathbf{r}}{a^3}  \bigg(\varphi^{q \ast} \varphi^{\ast} \varphi
  \varphi + \varphi^{\ast} \varphi^{q \ast} \varphi^q \varphi^q + \varphi^{\ast} \varphi^{\ast}
  \varphi \varphi^q + \varphi^{q \ast} \varphi^{q \ast} \varphi^q \varphi\nonumber\\
  & & - 2 \langle \varphi^{\ast} \varphi \rangle \varphi^{\ast} \varphi^q - 2
\langle \varphi^{\ast} \varphi \rangle \varphi^{q \ast} \varphi\bigg) \\
  & = &  S^{(2)}_I +  S^{(3)}_I +  S^{(4)}_I
 \end{eqnarray}

\begin{figure}[t]
    \centering
    \includegraphics[width=0.62\linewidth,]{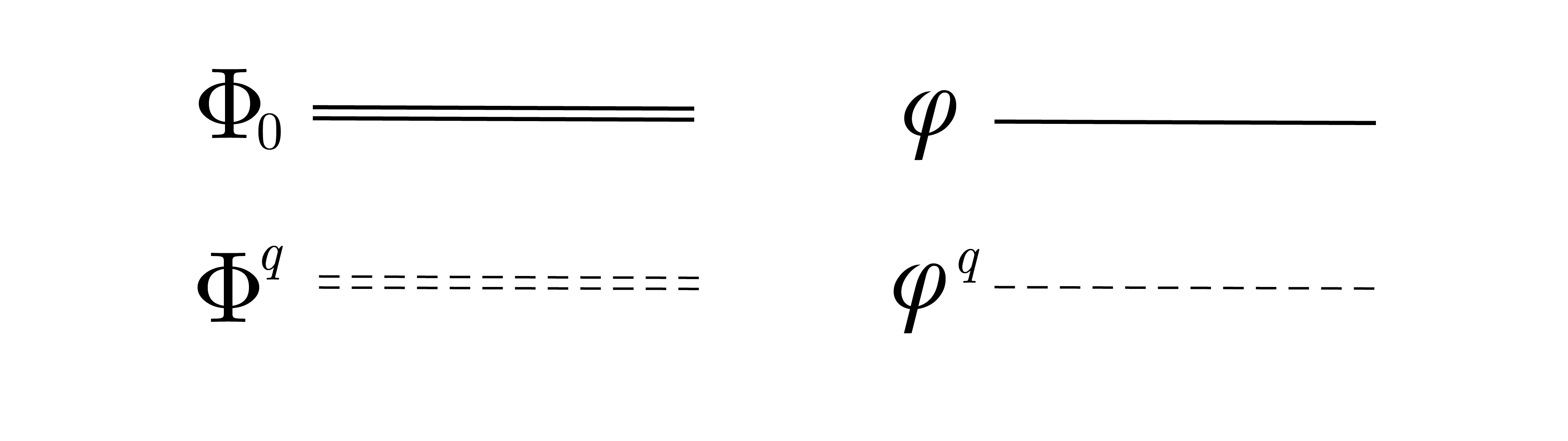}
    \caption{Notation for the lines representing the condensate (double lines) and the particles (single lines) in Feynman diagrams. Solid lines refer to classical fields while dashed lines to quantum fields.  
    }
    \label{fig:rules}
\end{figure}

The first piece $S_0$ defines the propagators used the Feynman rules while the second piece $S_I$ will provide the vertices needed for the construction of any propagator corrections due to the self-interaction or higher order correlators. Note that we have denoted three general types of interaction terms involving two, three and four $\varphi$ fields respectively. An important point that we establish as a rule is that since the $\Phi$ fields are slow, they will not be included inside a loop, but will just sit on external legs. In figure \ref{fig:rules} we show the notation used for the condensate and particle lines that we will use in the Feynman diagrams.

\subsection{Effective action for condensate and Schwinger - Dyson equations for particles}

\begin{figure}[t]
    \centering
    \includegraphics[width=0.4\linewidth,]{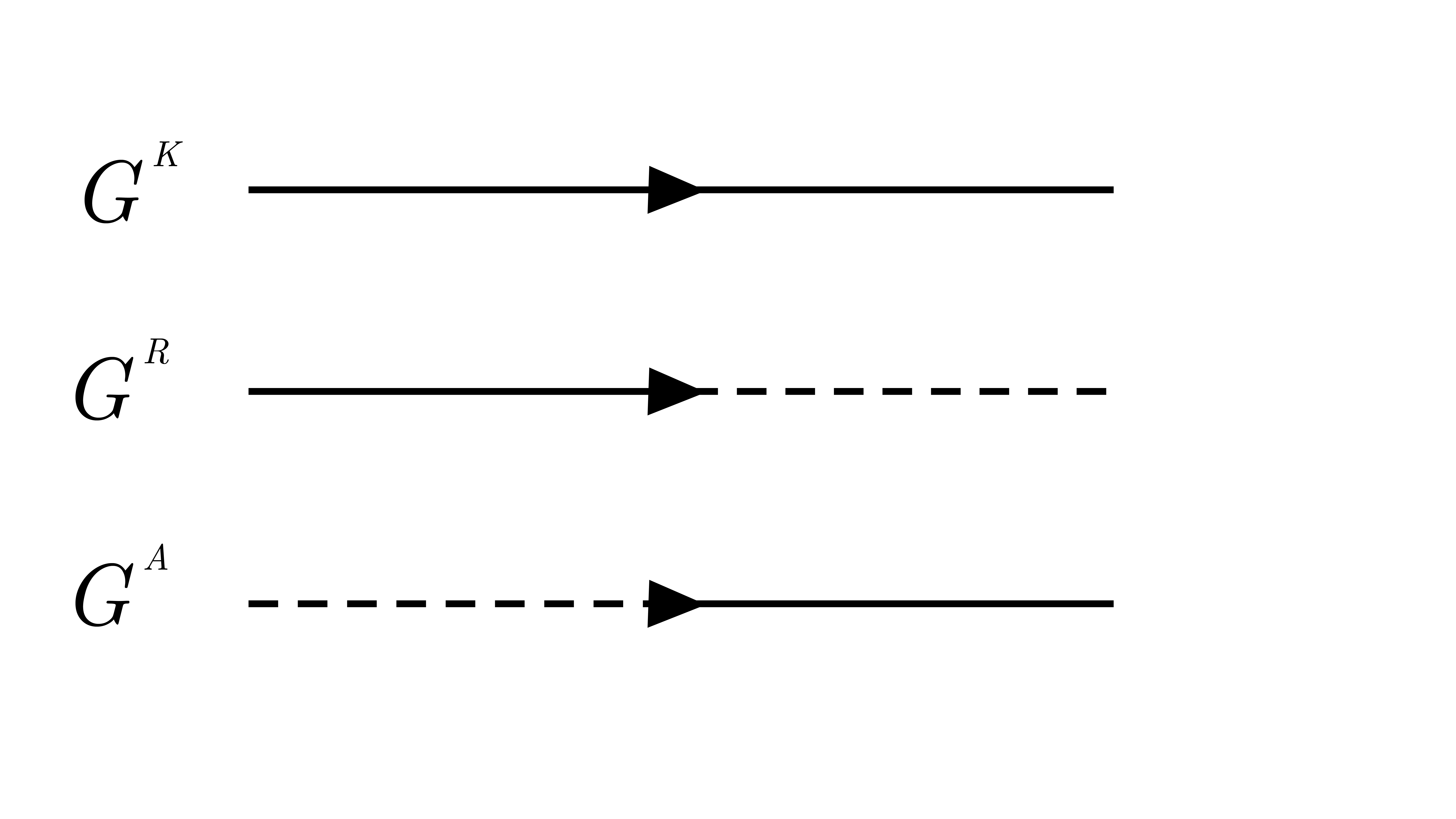}
    \caption{The three propagators for the $\varphi$ particles. $G^K$ is the $cl-cl$ propagator, $G^R$ the $cl-q$ (retarded) propagator and $G^A$ the $q-cl$ (advanced) propagator. The arrows point from the annihilation operator to the creation one, i.e.~from $\varphi$ to $\varphi^{\ast}$. 
    }
    \label{fig:propagators}
\end{figure}

We will firstly compute an approximation to the effective action for $\Phi$, taking into account the lowest order corrections coming from fluctuations of $\varphi$ fields. For this purpose we will use the generating functional
\begin{equation}
Z = \int \mathcal{D} [\Phi V \varphi] e^{i S_0 + i S_I}
\end{equation}
Expanding the interaction part and arranging, we have
\begin{equation}
Z = \int \mathcal{D} [\Phi V] e^{i S_{01}} \left( 1 + \frac{\int \mathcal{D}
[\varphi] e^{i S_{0 \varphi}} i S_I}{\int \mathcal{D} [\varphi] e^{i S_{0 \varphi}}} +
\frac{\int \mathcal{D} [\varphi] e^{i S_{0 \varphi}} \frac{1}{2} (i S_I)^2}{\int
\mathcal{D} [\varphi] e^{i S_{0 \varphi}}} + \cdots + \right) \int \mathcal{D}
[\varphi] e^{i S_{0 \varphi}}
\end{equation}
where $S_{01}$ is the quadratic action of the condensate and the gravitational potential, and $S_{0\varphi}$ the quadratic action of the particles. Integrating out the particles, we can write
\begin{equation}
Z = \int \mathcal{D} [\Phi V] e^{i S_{01}} \left( 1 + i \langle S_I \rangle -
\frac{1}{2} \langle S_I^2 \rangle + \cdots  \right) (\det (- i G^{- 1}_V))^{-
1}
\end{equation}
where
\begin{equation}
G_V^{-1}=G^{-1}_0-m V^q I
\end{equation}
with $I$ the identity matrix and
\begin{equation}
\label{g0}
G^{- 1}_0 = \left(\begin{array}{cc}
  0 & i \partial_t - H_{qp}\\
  i \partial_t - H_{qp} & 0
\end{array}\right)
\end{equation}
Due to our arrangement of terms with $\langle \varphi^{\ast} \varphi \rangle$ in the interaction part, it is easy to check that $\langle S_I \rangle = 0$. With this, and writing the determinant as an exponential, we can approximate the functional as
\begin{equation}
Z = \int \mathcal{D} [\Phi V] e^{i S_{01} - \frac{1}{2} \langle S_I^2 \rangle
- Tr (\ln (- i G^{- 1}_V))}
\end{equation}
Now, we consider the term $V^q$ as a perturbative parameter, so, up to first order in $V^q$:
\begin{equation}
\label{genfunceff}
Z = \int \mathcal{D} [\Phi V] e^{i S_{01} - \frac{1}{2} \langle S_I^2 \rangle
+ m V^q Tr(G_{0})}
\end{equation}
where $G_0$ is the inverse of $G^{-1}_0$ defined in \eqref{g0}. Its explicit form is 
\begin{equation}
G_0=\left(\begin{array}{cc}
  G_0^K & G_0^R\\
  G_0^A & 0
\end{array}\right)
\end{equation}
where $G^{R(A)}$ stands for the retarded (advanced) propagator and $G^K$ corresponds to the Keldysh propagator. Their representations in the Feynman rules are shown in figure \ref{fig:propagators}.

\begin{figure}[t]
    \centering
    \includegraphics[width=1\linewidth,]{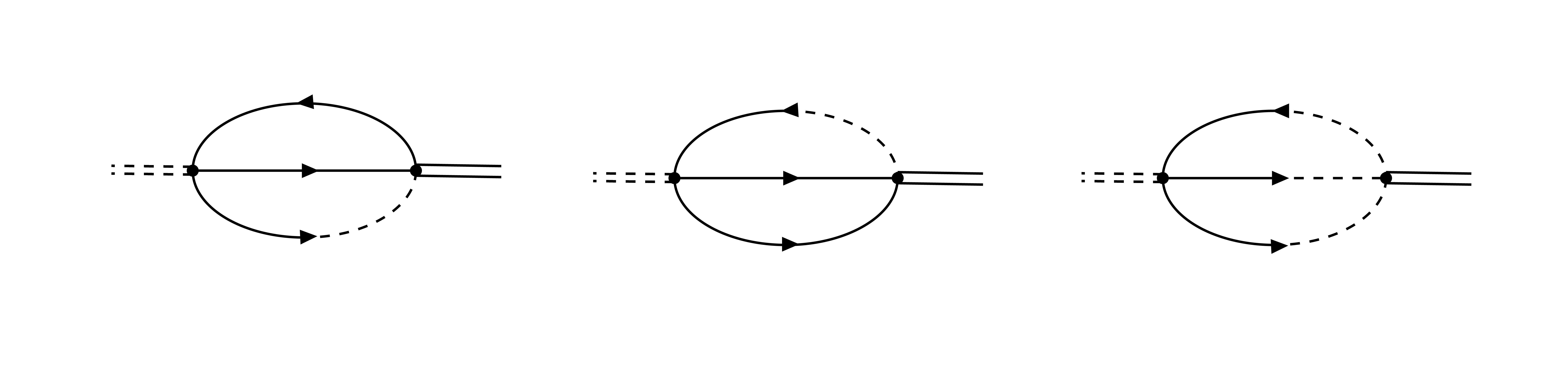}
    \caption{Feynman diagrams for the condensate retarded self-energy $\Sigma^R_{(cond)}$ due to $\varphi$ fluctuations at order $g^2$.} 
    \label{fig:sigmarcond}
\end{figure}

The term $\frac{i}{2} \langle S_I^2 \rangle$ corresponds to the order $g^2$ corrections to the condensate propagator due to fluctuations of the particles ($\varphi$ fields). Thus, this term contains the retarded (advanced) self-energies of the condensate, denoted by $\Sigma_{(cond)}^{R(A)}(x,x')$ - the diagrams of the retarded self-energies are shown in figure \ref{fig:sigmarcond}. Note that since we are working up to order 1 in $\Phi^q$ there is no Keldysh component of the self-energy here.
With this, from \eqref{genfunceff} we can define the effective action\footnote{Note that $\Sigma_{(cond)}(x,x')$ is a non-local 2-point object but we have suppressed the relevant second integration in (\ref{effaction1}) to avoid notational clutter.} 

\begin{eqnarray}
\label{effaction1}
 S_{eff} &  =  &\int d t d^3 \mathbf{r}
  \left(\begin{array}{cc}
    \Phi_0^{\ast} & \Phi^{q \ast}
  \end{array}\right) \left(\begin{array}{cc}
    0 & i \partial_t - H_c - \Sigma_{(cond)}^A\\
    i \partial_t - H_c - \Sigma_{(cond)}^R & 0
  \end{array}\right) \left(\begin{array}{c}
    \Phi_0\\
    \Phi^q
  \end{array}\right) \nonumber\\
& & + \frac{1}{4 \pi G} \int d t d^3\mathbf{r} \left(\begin{array}{cc}
    V^{cl} & V^q
  \end{array}\right) \left(\begin{array}{cc}
    0 & a\nabla^2\\
    a\nabla^2 & 0
  \end{array}\right) \left(\begin{array}{c}
    V^{cl}\\
    V^q
  \end{array}\right) \nonumber\\
& & - m \int d t d^3\mathbf{r} \, V^q (| \Phi_0 |^2 + i Tr(G_0))
\end{eqnarray}
Since $Tr(G_{0}) = G^K = - i \langle \varphi^{\ast} \varphi \rangle$
and here it is evaluated in one space-time point ($G^K (x, x)$ with $x = (t,
\mathbf{r})$), we can write the Keldysh propagator in terms of the
occupation number of the particles, that we will define as $f$, as $G^K (x, x) = - i \underset{k}{\sum} (2 f + 1)$ \cite{kamenev_2011}. We will neglect the last sum over $1$, since it gives a vacuum energy which can be
subtracted by means of a renormalisation. From this, we define the particle
number density as
\begin{equation}
\label{quasipartdens}
\tilde{n}=\underset{k}{\sum} f
\end{equation}
By the same token, we will write the condensate term $ | \Phi_0 |^2$ in terms of a number density. For this, we have to take into account that in the Schwinger-Keldysh formalism the normalization condition for the condensate is given by
\begin{equation}
\int d \mathbf{r} | \Phi_0 |^2 = 2 N_c
\end{equation}
where $N_c$ is the number of condensate particles, with this factor of 2 appearing due to the definition of the Keldysh rotation; moreover, we note that the expectation value for two classical fields is $\langle \phi^{cl} \phi^{cl \ast} \rangle \sim 2 n_B + 1$, with $n_B$ the total occupation number. It means that the condensate number density $n_c$ is given by
\begin{equation}
\label{condensatedens}
n_c = \frac{1}{2} | \Phi_0 |^2
\end{equation}
With these considerations, the effective action for the condensate can finally be written as
\begin{eqnarray}
\label{effaction2}
  S_{eff} &=& \int d t d^3\mathbf{r} \left(\begin{array}{cc}
    \Phi_0^{\ast} & \Phi^{q \ast}
  \end{array}\right) \left(\begin{array}{cc}
    0 & i \partial_t - H_c - \Sigma_{(cond)}^A\\
    i \partial_t - H_c - \Sigma_{(cond)}^R & 0
  \end{array}\right) \left(\begin{array}{c}
    \Phi_0\\
    \Phi^q
  \end{array}\right)\nonumber\\
  & & + \frac{1}{4 \pi G} \int d t d^3\mathbf{r} \left(\begin{array}{cc}
    V^{cl} & V^q
  \end{array}\right) \left(\begin{array}{cc}
    0 & a\nabla^2\\
    a\nabla^2 & 0
  \end{array}\right) \left(\begin{array}{c}
    V^{cl}\\
    V^q
  \end{array}\right)\nonumber\\
  & & - 2 m \int d t d^3\mathbf{r}\, V^q (n_c + \tilde{n})
\end{eqnarray}
with
\begin{eqnarray}
H_c &=& -\frac{1}{2 m a^2} \nabla^2 + m V^{cl} + \frac{g}{a^3}(n_c + 2\tilde{n})
\end{eqnarray}

The effective action \eqref{effaction2} provides the equations of motion for the condensate and the gravitational potential by varying with respect to the q-components. However, since the particles ($\varphi$ field) have been integrated out, their dynamics, and hence $\tilde{n}$, are not available from the above action. To determine the dynamical equations relevant to the non-condensed particles we now turn to a different route. For this, we can write down the Schwinger-Dyson equations for the particles' two point function, considering the corrections to the propagators coming from the vertices in $S_I$. Collecting the sum of all 1PI diagrams in the particle self-energy $\Sigma$, the series for the fully dressed particle propagator $G$ will schematically read like
\begin{equation}\label{SDeq1}
G = G_0+G_0 \otimes\Sigma\otimes G 
\end{equation}
if the series is factored from the left, or
\begin{equation}\label{SDeq2}
G = G_0+G \otimes\Sigma \otimes G_0
\end{equation}
if the series is factored from the right. Here, the product sign implies appropriate convolutions in spacetime and also multiplication in the Keldysh matrix space, noting in particular that the self energy $\Sigma$ takes the Keldysh form
\begin{equation}
    \Sigma  = \left(\begin{matrix}0 & \Sigma^A \\ \Sigma^R & \Sigma^K\end{matrix}\right)
\end{equation}
The left/right factorization is redundant in equilibrium, where convolutions are simply products in Fourier space. However, in a non-equilibrium situation the two forms contain different information \cite{RammerBook} and, as we shall see below, subtracting them leads to kinetic equation for the particles \cite{kamenev_2011}. Equation (\ref{SDeq1}) leads to 
\begin{eqnarray}
 \int d^4x' \left[(G_0^A)^{- 1}(x,x') - \Sigma^A(x,x')\right] G^A(x',y) = \delta (x - y) \label{eqga}\\
  \int d^4x' \left[(G_0^R)^{-1}(x,x') - \Sigma^R(x,x')\right] G^R(x',y) = \delta (x-y) \label{eqgr}\\
  \int d^4x' \left[(G_0^R)^{- 1}(x,x') - \Sigma^R(x,x')\right] G^K(x',y) = \int d^4x' \Sigma^K(x,x') G^A(x',y)  \label{eqgk}
\end{eqnarray}
where we denote $ x= (t, \mathbf{r})$ etc and we have written
\begin{equation}
(G_0^R)^{- 1}(x,x') = (G_0^A)^{- 1}(x,x') = \left[i
\partial_t + \frac{1}{2 m a^2} \nabla^2 - m V^{cl} - \frac{2g}{a^3}(n_c + \tilde{n})\right]\delta(x-x')
\end{equation}
where we have implicitly noted  the retarded and advanced initial conditions respectively for the definition of the operator. Equations \eqref{eqga} and \eqref{eqgr} will then give the explicit form of the advanced and retarded fully dressed propagators, $G^A(x,y)$ and $G^R(x,y)$ respectively. The self-energy $\Sigma(x,y)$, when computed in some approximation, provides via the above the dressed propagator of the particles described by the $\varphi$ field.  Summarising the above quantities, $\Sigma^{R (A)}$ corresponds to the retarded (advanced) self-energies for
the particles, $\Sigma_{(cond)}^{R (A)}$ stands for the retarded
(advanced) self-energies for the condensate, $n_c$ is the condensate number density and $\tilde{n}$ the particle number density. It is important to recall that
these self-energies are order $g^2$ or higher, since the order $g$ was
introduced in $H_c$ and $H_{q p}$ because we have introduced first the slow
varying quantities in the matrices as part of an effective potential.

\section{General equations for bosonic Dark Matter} 

We now use the results of the previous section to obtain the combined equations for condensed and non-condensed scalar dark matter.  

\subsection{Kinetic equation for the particles}
The Keldysh part of the full Schwinger-Dyson equation, Eq.~\eqref{eqgk}, can be used to provide the kinetic equation for the particle component \cite{1988PhRvD..37.2878C, RammerBook, kamenev_2011}. Following in particular the analysis of \cite{kamenev_2011}, we first parameterize the Keldysh correlator using a hermitian matrix $F$ as $G^K = G^R F - F G^A$, where $F(x,y)$ is a distribution function depending on two space-time points. This distribution $F$ can be related with the occupation number of the particles $f$, as 
\begin{equation}
\label{relationf}
F=2f+1.
\end{equation}
By multiplying \eqref{eqgk} from the right by $ ((G_0^A)^{- 1} - \Sigma^A)$, using \eqref{eqga} and \eqref{eqgr} and after some algebra, we arrive at
\begin{equation}
\label{eqpropgk}
  - \left[ i \partial_t + \frac{1}{2 m a^2} \nabla^2 - V_{eff} (y), F
  \right] = \Sigma^K - [\Sigma^R, F] - F (\Sigma^R - \Sigma^A)
\end{equation}
where the l.h.s. involves the commutator and we have defined
\begin{equation}
  V_{eff} = m V^{cl} + \frac{2g}{a^3}(n_c + \tilde{n})
\end{equation}

Now, we will assume that the fast modes' intrinsic scales of spatiotemporal variation are much smaller than both galactic length scales and the typical evolutionary times we are considering. To take advantage of this, we will employ the Wigner transform which for an arbitrary function $A(y_1,y_2)$ corresponds to a change of variables to the central point $y=\frac{1}{2}(y_1+y_2)$ and the relative coordinate $y'=y_1-y_2$, and then a Fourier transform of the dependence on $y'$. For functions of slow variation in the central coordinate $y$ and fast variation in the relative coordinate $y'$, the Wigner transform of products of functions, and therefore also the commutator, becomes much easier to treat algebraically.

With these considerations in mind we take a Wigner transform of \eqref{eqpropgk}. For this, when we have to compute a product of two Wigner transforms, we use the fact that as $V^{cl}$ and $\Phi_0$ are slow varying fields, and one can assume that both number densities vary slowly in time. Also, we will work in a regime where we consider the scale factor $a(t)$ as also slowly varying. Hence, after some algebra and with \eqref{relationf}, we  obtain
\begin{eqnarray}
\label{kineticeq}
& & \left( 1 - \frac{\partial}{\partial \varepsilon} \Re (\Sigma^R)
  \right) \frac{\partial}{\partial t} f(t,\mathbf{r},\mathbf{k}) + \left( \frac{k}{m a^2} +
  \nabla_{\mathbf{k}} \Re (\Sigma^R) \right)
  \nabla f(t,\mathbf{r},\mathbf{k}) \nonumber\\
& & - \nabla (V_{eff}
  + \Re (\Sigma^R)) \nabla_{\mathbf{k}} f(t,\mathbf{r},\mathbf{k}) = \frac{1}{2}I^{coll} [f]
\end{eqnarray}
where
\begin{equation}
\label{collisional}
  I^{coll} [f] = i \Sigma^K + 2 (2f+1) \Im (\Sigma^R)
\end{equation}
and here $t$, $\mathbf{r}$ corresponds to the central point coordinates,  $\mathbf{k}$ is the momentum related with the Fourier transform of the relative coordinate, and $\Re$, $\Im$ respectively denote real and imaginary parts. We observe that Eq.~(\ref{kineticeq}) corresponds to a Boltzmann-like equation.

\subsection{Main equations}

We are now in a position to present the main equations of our model. We get the equations of motion for the condensate and the gravitational potential by varying the action \eqref{effaction2} with respect to $\Phi^{q \ast}$ and $V^q$. On the other hand, for the particles we use equation \eqref{kineticeq} considering that the effective potential variation in time is slow and we neglect the real part of the self-energies, which just contributes to small shifts in \eqref{kineticeq}. 
With all of these elements, the equations for the condensate, particles and gravitational potential read
\begin{eqnarray}
 && i \partial_t \Phi_0 = \bigg[- \frac{1}{2 m a^2} \nabla^2 + (m V + \frac{g}{a^3} (n_c + 2
  \tilde{n}))-i R \bigg] \Phi_0 \label{eq:final BEC}\\
&&  \frac{\partial f}{\partial t} + \frac{k}{m a^2} \nabla f
  - \nabla \bigg(m V +  \frac{2g}{a^3} (n_c + \tilde{n})\bigg)
  \nabla_{\mathbf{k}}f  = \frac{1}{2}(I_a + I_b)\\
&&  \nabla^2 V = \frac{4 \pi G m}{a} (n_c + \tilde{n})
\end{eqnarray}

where
\begin{eqnarray}
I_{a} &=& 8 \pi \frac{g^2}{a^6} n_c \mathbf{\underset{\mathbf{p},
\mathbf{l}, k}{\sum}} \delta (\varepsilon_{\mathbf{q}} +
\varepsilon_{\mathbf{p}} - \varepsilon_{\mathbf{k}} -
\varepsilon_{\mathbf{l}}) \delta (\mathbf{l} - \mathbf{p} - \mathbf{q}
+ \mathbf{k}) \bigg( \delta (\mathbf{p} - \mathbf{r}) - \delta (\mathbf{k} -
\mathbf{r}) \nonumber\\ 
& & - \delta (\mathbf{l} - \mathbf{r}) \bigg) 
\bigg[(1 + f_p) f_k f_l - f_p (1 + f_k) (1 + f_l)\bigg] \label{colla}\\
I_{b} &=& 8 \pi \frac{g^2}{a^6} \underset{\mathbf{p}, \mathbf{q}, \mathbf{l}}{\sum}
\delta (\varepsilon_{\mathbf{p}} + \varepsilon_{\mathbf{q}} -
\varepsilon_{\mathbf{k}} - \varepsilon_{\mathbf{l}}) \delta (\mathbf{l}
- \mathbf{p} - \mathbf{q} + \mathbf{k}) \nonumber\\
& & \times \bigg[ f_p f_q (f_k + 1) (f_l + 1) -
f_k f_l (f_p + 1) (f_q + 1) \bigg]\label{collb}
\end{eqnarray}
and 
\begin{equation}\label{eq:finalR}
R = \frac{1}{4 n_c} \underset{\mathbf{r}}{\sum} I_a
\end{equation}
In the equations for $I_a$ and $I_b$ we have used the shorthand $f_p$ for $f(t,r,\mathbf{p})$ and $\varepsilon_\mathbf{p}$ for the energy function $\varepsilon(\mathbf{r},\mathbf{p})$.
We note that the term $I_b$ corresponds to collisions between particles and is number-preserving. On the other hand, the collisional term $I_a$ represents the process of particles scattered into and out of the condensate. The $R$ term reflects this exchange in the condensate equation, driving to a growth or a decrease of the condensate.
Nonetheless, we note that the total particle number is conserved within our formalism.

Equations \eqref{eq:final BEC} - \eqref{eq:finalR} are the main results of this paper. The details of the derivation of the self-energies needed to get these equations and the collisional terms $I_a$ and $I_b$ are described in appendix \ref{appB}.

\subsection{Limits and implications of the main equations}

Some comments on the above equations are in order. Our derivation was rather general, assuming only a separation of spatiotemporal scales. Therefore, the dynamical equations encompass all known models involving the  underlying bosonic field action used here and, in appropriate limits, reduce to the models that have been used in the literature, as briefly demonstrated below.

\subsubsection{Fuzzy Dark Matter}

Firstly, let us to consider the limit where there are no self-interactions ($g=0$). In this case there are no collisional or damping terms ($I_a = I_b = R = 0$). Also, if we consider the condensate to be the dominant contribution, neglecting the higher momentum particles, we recover the Schr\"{o}dinger-Poisson equations used in Fuzzy Dark Matter
\begin{eqnarray}
  i \partial_t \Phi_0 &=& \bigg[- \frac{1}{2 m a^2} \nabla^2 + m V\bigg] \Phi_0\\
  \nabla^2 V &=& \frac{4 \pi G m}{a} n_c
\end{eqnarray}
This set of equations have been extensively analysed \cite{2016PhR...643....1M} and, for example, simulated in \cite{2014NatPh..10..496S, 2016PhRvD..94d3513S, 2017MNRAS.471.4559M, Veltmaat:2018dfz, May_2021, Nori:2022afw}.

\subsubsection{Self-Interacting Fuzzy Dark Matter}
Now, if we allow $g\neq 0$ but we continue ignoring the particles the equations become 
\begin{eqnarray}
  i \partial_t \Phi_0 &=& \bigg[- \frac{1}{2 m a^2} \nabla^2 + m V + \frac{g}{a^3} n_c \bigg] \Phi_0\\
  \nabla^2 V &=& \frac{4 \pi G m}{a} n_c
\end{eqnarray}
which corresponds to the situation of Self-Interacting Fuzzy Dark Matter, which has been studied e.g.~in \cite{Boehmer:2007um, Chavanis_2011,PhysRevD.84.043532,Rindler-Daller:2011afd, Guth:2014hsa, Chen:2021oot, Kirkpatrick:2021wwz, Hartman:2022cfh, Delgado:2022vnt, Mocz:2023adf}. Note that the importance of even a small self-interaction for an ultralight axionic field was stressed in \cite{PhysRevD.97.023529}.

\subsubsection{Vlasov-Poisson CDM}
On the other hand, if the high momentum particles are the dominant component and we can neglect the existence of the condensate together with $g=0$, we recover the Vlasov-Poisson equations used to describe Cold Dark Matter
\begin{eqnarray}
& & \frac{\partial f}{\partial t} + \frac{k}{m a^2} \nabla f - m\nabla V
  \nabla_{\mathbf{k}}f = 0  \\
& & \nabla^2 V = \frac{4 \pi G m}{a} \tilde{n}
\end{eqnarray}
which are typically used for particles with much higher masses compared with those used in Fuzzy Dark Matter, along with the assumption that dark matter is weakly interacting. The distribution $f$ is what is sampled in N-body solvers \cite{Bernardeau+, Angulo+Hahn}.

\subsubsection{Boltzmann-Poisson DM}
The simplest extension to Vlasov-Poisson is a Boltzmann-Poisson equation, in which the condensate component is again set to zero, but the collisional term $I_b$ is explicitly included 
\begin{eqnarray}
& & \frac{\partial f}{\partial t} + \frac{k}{m a^2} \nabla f - \nabla \left(mV+\frac{2g}{a^3}\tilde{n}\right)
  \nabla_{\mathbf{k}}f = \frac{1}{2} I_b  \\
& & \nabla^2 V = \frac{4 \pi G m}{a} \tilde{n}
\end{eqnarray}
Note the appearance of a term linear in the self-coupling $g$, in addition to the collision integral $I_b$ which is quadratic in $g$. Such a model could be treated via a fluid approximation under the assumption of interactions strong enough to establish local thermodynamic equilibrium over short time scales.

\subsubsection{ZNG Formalism}
The generality of our formalism is further evidenced by the fact that it completely encompasses the so-called `Zaremba-Nikuni-Griffin' (or ZNG) formalism used in ultracold atomic gases~\cite{zaremba1999dynamics,griffin_nikuni_zaremba_2009,proukakis2008finite}.
To see this we consider the simplified limit -- not relevant to the cosmological setting -- when we ignore all gravitational (interaction) effects, and also the effect of cosmic expansion, done by setting
 $a=1$. The equations then become
\begin{eqnarray}
  i \partial_t \Phi_0 = \bigg[- \frac{1}{2 m} \nabla^2 +  g(n_c + 2
  \tilde{n})-i R \bigg] \Phi_0\\
  \frac{\partial f}{\partial t} + \frac{k}{m} \nabla f
  - 2g\nabla \bigg(n_c + \tilde{n}\bigg)
  \nabla_{\mathbf{k}}f = \frac{1}{2}(I_a + I_b) \;.
\end{eqnarray}
This is precisely the set of a coupled dissipative Gross-Pitaevskii equation and a collisional quantum Boltzmann equation used within the ZNG formalism of cold atomic physics to describe the coupled dynamics of a finite temperature atomic condensate co-existing with a thermal cloud \cite{zaremba1999dynamics,griffin_nikuni_zaremba_2009}.
Such an approach has been used to successfully describe a range of experimental observations, including damping of various collective modes~\cite{Jackson-ZNG-Collective,griffin_nikuni_zaremba_2009}, soliton decay~\cite{Jackson-ZNG-soliton-PhysRevA.75.051601}, vortex decay~\cite{Jackson2009-ZNG}, Josephson effects~\cite{Xhani20-ZNG}, condensate growth dynamics (upon imposing a small numerical seed to set off the calculations)~\cite{zaremba-stoof-growth-PhysRevA.62.063609,Allen-ZNG-evap}, and even condensate mixtures~\cite{Lee_2016-ZNG-review,Lee_2018-ZNG}.

We therefore see that, in addition to the various limiting cases exposed above, our description also  reproduces correctly a physical system, namely a cold atomic dilute gas, where there is a known and extensively studied interplay between the condensed and the incoherent particle components of the underlying field.

\begin{center}
    $\star\star\star$ 
\end{center}

We can therefore conclude that the model described by our equations generalises current perspectives on the theoretical description of DM, simultaneously including various features in a general manner and thus containing both CDM and the Fuzzy Dark Matter model as its limiting cases. Furthermore, our model is more general than any of them due to the presence of self-interactions. Moreover, we note that if one were to follow our procedure but only keep terms just up to order $g$, one would then arrive at the reduced equations
\begin{eqnarray}
  && i \partial_t \Phi_0 = \bigg[- \frac{1}{2 m a^2} \nabla^2 + \left(m V + \frac{g}{a^3} (n_c + 2
  \tilde{n})\right) \bigg] \Phi_0\\
  && \frac{\partial f}{\partial t} + \frac{k}{m a^2} \nabla f
  - \nabla \left(m V +  \frac{2g}{a^3} (n_c + \tilde{n})\right)
  \nabla_{\mathbf{k}}f = 0 \\
  && \nabla^2 V = \frac{4 \pi G m}{a} (n_c + \tilde{n}) \;.
\end{eqnarray}
This demonstrates explicitly that, even if collisional terms are ignored, the coherent and incoherent components of the system are coupled non-trivially, with condensate and particles feeling the effect of each other (through mean-field coupling only in such limit). It is also worth noting that, consistent with the usual Hartree-Fock considerations, their effective potentials are slightly different with different prefactors for $n_c$ and $\tilde{n}$ in the two equations above. Collisional terms only appear on the RHS of the Boltzmann equation when considering terms up to order $g^2$. As a result, in the absence of collisions, not only is the total number of particles  conserved, but also the number of particles in each of the condensate and non-condensate components is individually preserved.~\footnote{A mixture of CDM with condensate dark matter has been discussed in \cite{Vogt:2022bwy}, where they study such mixing through a Halo Model. Here, we are considering a completely different model. The two components in our equations come from the same initial action. Thus, our non-condensed particles are of the same nature as those of the condensate, only belonging to a different physical phase.}

\section{Final remarks}

In this work we have charted a path from a non-relativistic, bosonic action to a set of dynamical equations for a generic dark matter model involving a scalar particle with mass $m$ and quartic self-interaction with self-coupling strength $g$. Starting from the fundamental, non-equilibrium partition function, our final equations describe states where the dark matter can contain both a coherent (or condensed) component governed by a wave equation, as well as an incoherent  component corresponding to a collection of particles described by a (quantum) Boltzmann kinetic equation for their phase space density. Such Boltzmann equation includes collisional terms that arise at order $g^2$ in the self-interaction, and thus facilitates the transfer of particles between the coherent/condensate and incoherent/particle components. This description thus encompasses a variety of dark matter models discussed in the literature, including both corpuscular dark matter (like CDM), and wave dark matter (like Fuzzy Dark Matter or $\psi$DM) under a unified framework. Note that both types of dark matter are described by the same underlying scalar field and the existence of the two components depends on the model parameters and the available energy.  
Such a picture is directly analogous to the mature, and experimentally-confirmed, study of coherent and incoherent components in atomic/condensed-matter systems when the dimensionless phase-space density $(\rho/m)\lambda_{\rm dB}^3 \gtrsim O(1)$, which typically arises (in the context of ultracold atomic gases, or liquid helium) below some characteristic threshold (or critical) temperature: in the latter context, the system is typically described by some appropriately generalized (depending on the approximations made within different models) version of the Gross-Pitaevskii equation for the coherent (typically low-momentum) part, self-consistently coupled to a quantum Boltzmann equation for the incoherent (typically higher momentum) part of the system, and such mature description has inspired the presented study.

The description of this fundamental cosmological bosonic system has attracted a lot of attention in the literature as an alternative to cold dark matter, which, due to its bosonic nature, can exhibit condensation and related wavy features given the right circumstances. Beyond discussions on structure formation and the resulting cored density profile of dark matter halos, attention has also been drawn more generally to the notion of Bose-Einstein Condensation on cosmological scales and its possible meaning. An initial suggestion by \cite{Sikivie:2009qn} of gravitational thermalization and the formation of condensed states with enormous correlation lengths has been critically investigated in \cite{Guth:2014hsa}. The latter work noted that gravitationally induced condensation does indeed occur, as suggested by \cite{Sikivie:2009qn}, but found that the equilibrium state of light particles interacting through gravity (and more generally attractive interactions) must be made up of so called Bose stars (solitons), engulfed in an incoherent field, thus inhibiting the development of very large correlation lengths. The formation of Bose stars via gravitational condensation from a random initial condition has been verified via the work of \cite{Levkov:2018kau} and a detailed study of the coherence of gravitationally bound bosonic halos \cite{2022arXiv221102565L} seems to agree with the suggestion of \cite{Guth:2014hsa} for an inhibited correlation length of the equilibrium state. Further work on the role of gravitation and self-interactions in condensation has been pursued in \cite{Levkov:2018kau, Chen:2021oot, Kirkpatrick:2021wwz, Bar-Or:2018pxz, Chavanis:2020upb} where kinetic equations derived from a wave equation have been used, somewhat analogously to the study of wave turbulence based on a Boltzmann equation, and is expected to describe well certain dynamical regimes. Our approach explicitly maintains all collisional factors in a more general (quantum) Boltzmann equation, which is itself coupled to the coherent degrees of freedom of the system, thus providing a more extended description of the coupling of coherent and incoherent parts of the system. This explicit combination of a wave equation, self-consistently coupled to a (quantum) Boltzmann equation thus facilitates consideration of both weak wave turbulence and strong turbulence, in the sense of also facilitating emerging quantum vortex structures. It would be interesting to compare and contrast such an approach to the formalism derived here, to delineate more clearly the regimes of cosmological interest where their predictions might differ.

A few final comments are in order. The implications of a condensate-particle mixture for the early, linear stages of cosmological structure formation are left to be explored and will be the focus of upcoming work, including the effects of a self-coupling. Furthermore, the general conditions under which such a coupled mixture offers a useful description of bosonic dark mater remain to be clarified. For example, when solitonic cores have formed in the centres of halos, the condensed component is highly dominant there and likely to be the only relevant dynamical entity, irrespective of the coupling to the non-condensed component. However, as demonstrated in \cite{2022arXiv221102565L}, a transition region exists between the core and the outer halo where both components will be of similar density and hence dynamically relevant and coupled. And of course the relation between the possible description of the turbulent outer halo via the Boltzmann equation still needs to be clarified  - see the discussion in the above paragraph in relation to other works on kinetic descriptions. Finally, we point out that the equations presented in this paper reflect the leading order in a semiclassical expansion of the fundamental partition function \eqref{eq:PartitionFn} for the condensate component. Computation to the next order in $\Phi^q$ and $V^q$ reveals \emph{stochastic forces} that act on the condensate like those described in \cite{stoof1999coherent} for the case of a trapped atomic gas; the relevant derivation of these extended equations is the subject of a forthcoming publication. Such stochastic equations may be relevant to the study of the spontaneous soliton formation described in \cite{Levkov:2018kau, Chen:2021oot}.

We close by noting that the starting point of this work, namely non-equilibrium QFT in its Schwinger-Keldysh incarnation, has been extensively used to address various aspects of axion-like particle dynamics, see e.g.~\cite{Cao:2022bua} for a recent example. We hope to be able to explore connections with the formalism developed here in the near future. Furthermore, our approach may be useful for studying alternative dark matter models where the underlying bosonic dynamics is different from (\ref{eqini}) and where an interplay of condensate dynamics and a coupling to baryonic matter is relevant, as in \cite{Berezhiani:2015bqa}.

\section*{Acknowledgements}
This work was supported by the Leverhulme Trust, Grant no.
RPG-2021-010. We acknowledge extensive collaboration on numerical FDM modelling with Dr Gary (I-Kang) Liu, who also performed the FDM simulations and prepared the image shown in Fig.~1(a).

\appendix

\section{Diagonalisation and Popov approximation}\label{appA}

Let us consider the particle quadratic part in the action \eqref{action1}, given by  
\begin{equation}
S_\xi=\frac{1}{2} \int d t d^3\mathbf{r} \left(\begin{array}{cc}
  \xi^{\dag} & \xi^{q \dag}
\end{array}\right) \left(\begin{array}{cc}
  0 & M\\
  M & 0
\end{array}\right) \left(\begin{array}{c}
  \xi\\
  \xi^q
\end{array}\right)
\end{equation}
To obtain the particle energy spectrum and simplify the computations, $M$ should be put in a diagonal form. For this purpose we will use the Bogoliubov transformation:
\begin{eqnarray}
\varphi (\mathbf{r}, t) &=& \underset{k}{\sum} (u_k (\mathbf{r}) \eta_k (t)
+ v_k^{\ast} (\mathbf{r}) \eta_k^{\ast} (t))\\
\varphi^q (\mathbf{r}, t) &=& \underset{k}{\sum} (u_k (\mathbf{r}) \eta^q_k
(t) + v_k^{\ast} (\mathbf{r}) \eta_k^{q \ast} (t))
\end{eqnarray}
where the mode functions $u_k(\mathbf{r})$ and $v_k(\mathbf{r})$ are determined via the solution of the Bogoliubov-de Gennes equation
\begin{equation}
   \left(\begin{array}{cc}
    H_{qp} &  \frac{g}{2 a^3} \Phi_0 \Phi_0\\
   \frac{g}{2 a^3} \Phi^{\ast}_0 \Phi^{\ast}_0 &  H_{qp}
\end{array}\right) \left(\begin{array}{c}  u_k(\mathbf{r})\\v_k(\mathbf{r})
\end{array}\right) = \varepsilon_k' \left(\begin{array}{c}  u_k(\mathbf{r})\\ - v_k(\mathbf{r})
\end{array}\right)
\end{equation}
and are normalized according to
\begin{equation}
\int d^3 \mathbf{r} (u^{\ast}_k (\mathbf{r}) u_l (\mathbf{r}) -
v_k^{\ast} (\mathbf{r}) v_l (\mathbf{r})) = \delta_{k l}
\end{equation}
The label $k$ refers to the the compete set of solutions of the above eigenfunction equation.

In terms of the variable $\xi$, the transformation is defined as
\begin{equation}
\xi (\mathbf{r}, t) = \underset{k}{\sum} U_k (\mathbf{r}) \varsigma_k
(t), \qquad \xi^q (\mathbf{r}, t) = \underset{k}{\sum} U_k (\mathbf{r})
\varsigma_k^q (t)
\end{equation}

with $\varsigma (t) = \left(\begin{array}{c}
  \eta_k (t)\\
  \eta_k^{\ast} (t)
\end{array}\right)$, $\varsigma^q (t) = \left(\begin{array}{c}
  \eta^q_k (t)\\
  \eta_k^{q \ast} (t)
\end{array}\right)$ and $U (\mathbf{r}) = \left(\begin{array}{cc}
  u_k (\mathbf{r}) & v^{\ast}_k (\mathbf{r})\\
  v_k (\mathbf{r}) & u_k^{\ast} (\mathbf{r})
\end{array}\right)$.
With this, the quadratic part reads
\begin{equation}
S_\xi=\frac{1}{2}\int d t d^3\mathbf{r} \underset{k}{\sum}
(\varsigma_k^{\dag} (t) U_k^{\dag} (\mathbf{r}) M U_k (\mathbf{r})
\varsigma_k^q (t) + \varsigma_k^{q \dag} (t) U_k^{\dag} (\mathbf{r}) M U_k
(\mathbf{r}) \varsigma_k (t))
\end{equation}
This equation can be written in terms of a diagonal matrix $\tilde{M}$ as
\begin{equation}
S_\xi=\frac{1}{2} \int d t \underset{k}{\sum} (\varsigma_k^{\dag} (t) \tilde{M}_k
\varsigma_k^q (t) + \varsigma_k^{q \dag} (t) \tilde{M}_k \varsigma_k (t))
\end{equation}
with $\tilde{M}_k = \left(\begin{array}{cc}
  i \partial_t - \varepsilon'_k & 0\\
  0 & - i \partial_t - \varepsilon'_k
\end{array}\right)$ if the following relation is satisfied:
\begin{equation}
\varepsilon'_k = \frac{(| u_k |^2 - | v_k |^2)^2}{| u_k |^2 + | v_k |^2}
\varepsilon_k
\end{equation}
where $\varepsilon_k$ is an energy eigenvalue of the Hamiltonian $H_{qp}$ defined in $M$.

We will make the approximation that the diagonalization process doesn't change the quasiparticle energy spectrum very much, $\varepsilon'_k \approx \varepsilon_k $. This means that we can approximate $| u_k |^2 \approx 1$ and $| v_k
|^2 \approx 0$. In practical terms we are just ignoring the off-diagonal entries of $M$ and keeping approximately the same hamiltonian $H_{qp}$ with the same field $\varphi(\mathbf{r},t)$ in our action. We observe that with this approximation
\begin{equation}
\left(\begin{array}{cc}
  0 & \tilde{M}\\
  \tilde{M} & 0
\end{array}\right) \approx \left(\begin{array}{cccc}
  0 & 0 & i \partial_t - H_{qp} & 0\\
  0 & 0 & 0 & - i \partial_t - H_{qp}\\
  i \partial_t - H_{qp} & 0 & 0 & 0\\
  0 & - i \partial_t - H_{qp} & 0 & 0
\end{array}\right)
\end{equation}
The inverse of this matrix is such that the elements $\langle \varphi
\varphi \rangle$, $\langle \varphi^{\ast} \varphi^{\ast} \rangle$, $\langle
\varphi^q \varphi^q \rangle$ and $\langle \varphi^{q \ast} \varphi^{q \ast}
\rangle$ are zero, which is the Popov approximation often used in kinetic models in cold atomic systems~\cite{griffin-1996,proukakis2008finite,griffin_nikuni_zaremba_2009}. Thus, this diagonalization and the assumption $H'_{qp} \approx \bar{H}_{qp}$ ($|
u_k |^2 \approx 1$ and $| v_k |^2 \approx 0$) is equivalent to making the Popov approximation.

\section{Computation of the order $g^2$ collision terms}\label{appB}

We will compute the equations of the model up to order $g^2$ which introduces non-zero collisional terms in \eqref{collisional} and a modification to the condensate equation. Recall that $\Sigma^R$ and $\Sigma^K$ start at order $g^2$ since the order $g$ was included in the effective potential of the hamiltonian. We will first compute the self- energies which finally lead to the collisional terms.

\subsection{Condensate self-energy $\Sigma_{(cond)}^R$}

We compute the diagrams that contributes to $\Sigma_{(cond)}^R$ - see figure \ref{fig:sigmarcond}. Putting them together, we have that
\begin{eqnarray}
- i \Sigma_{(cond)}^R &=& - \frac{g^2}{8 a^6} \bigg[ 4 i^3 G^R (y, y') G^K
(y, y') G^K (y', y) + 2 i^3 G^A \left( y', y \right) G^K (y, y')^2 \nonumber\\
& & + 2 i^3 G^A (y', y) G^R (y, y')^2 \bigg]
\end{eqnarray}
We can replace the two $G^R (x, x')$ in the last term by $(G^R
(x, x') - G^A (x, x'))$, since $G^A (x', x) G^A (x, x') = 0$. We use the Wigner transform, noting that we work in a regime where the scale factor $a$ varies slowly. Furthermore, we define $(G^R (p) - G^A (p)) = - 2 \pi i \delta (\varepsilon - \varepsilon_{\mathbf{p}})$ and $G^K (p) = - 2 \pi i
F (p) \delta (\varepsilon - \varepsilon_{\mathbf{p}})$. Also, we symmetrize the first term between $k$ and $p+q-k$ and then we take the imaginary part. The real part is just a renormalization to the particle density; we will neglect the effects of the real parts in the retarded self-energies. We then perform the energy integration and we get that the imaginary part of the condensate self-energy $\Sigma_{(cond)}^R$ is given by
\begin{eqnarray}
\label{sigmacondRfin}
\Im (\Sigma_{(cond)}^R) &=& - \frac{\pi g^2}{2 a^6} \underset{\mathbf{p},
\mathbf{k}}{\sum} \delta (\varepsilon_{\mathbf{q}} +
\varepsilon_{\mathbf{p}} - \varepsilon_{\mathbf{k}} -
\varepsilon_{\mathbf{p} + \mathbf{q} - \mathbf{k}}) \bigg[F (p) \bigg(F (k)+ F (p + q - k)\bigg)  \nonumber\\
& & - (F (k) F (p + q - k) + 1)\bigg]
\end{eqnarray}

\subsection{Particle self-energy $\Sigma^R$}

\begin{figure}[t]
    \centering
    \includegraphics[width=1\linewidth,]{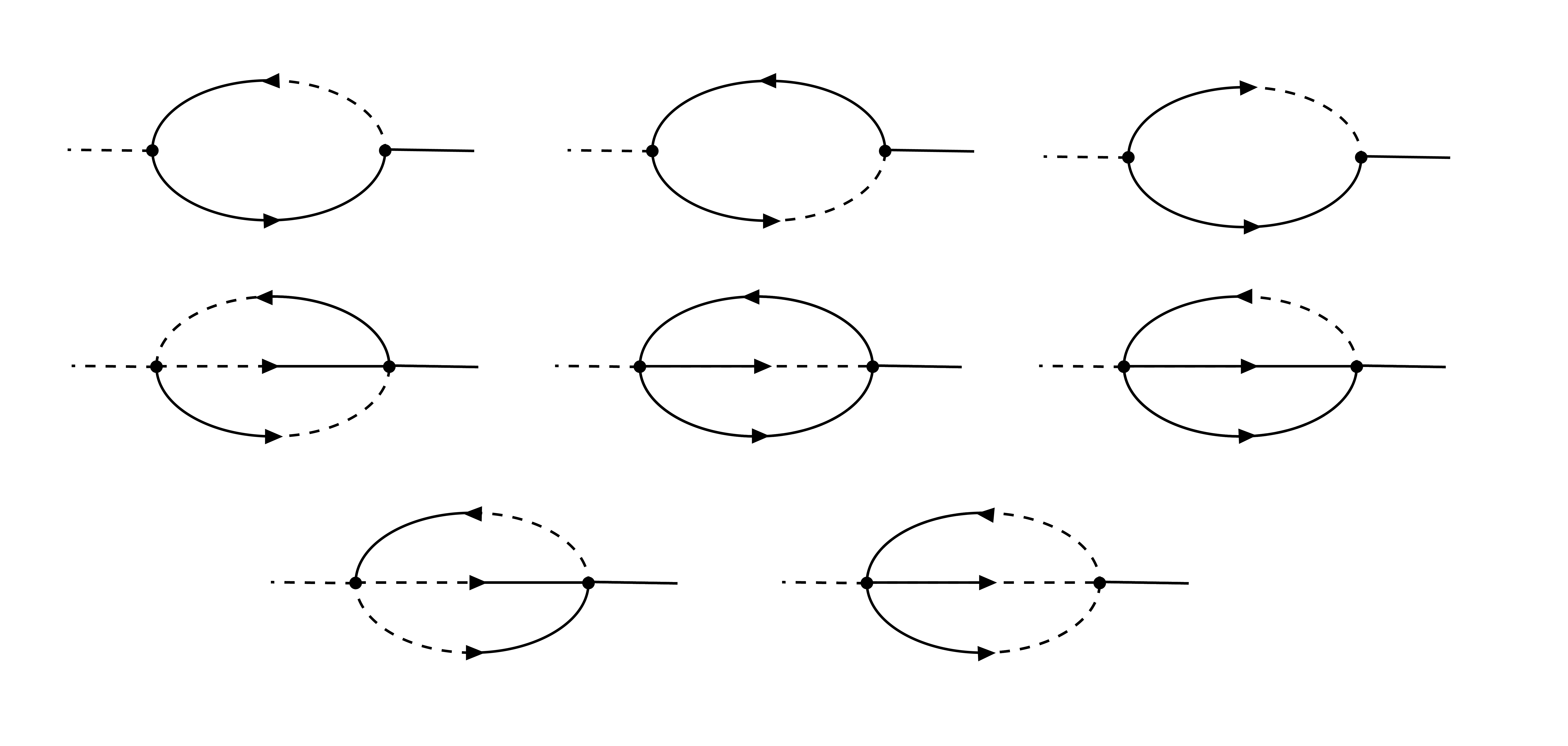}
    \caption{Feynman diagrams for the particle retarded self-energy at order $g^2$. The three upper diagrams correspond to $\Sigma^{R}_{(a)}$, the remaining corresponds to $\Sigma^{R}_{(b)}$.
    }
    \label{fig:sigmarpart}
\end{figure}

The diagrams that contribute to the particle retarded self-energy $\Sigma^R$ are presented in figure \ref{fig:sigmarpart}. 
We write them as
\begin{equation}
- i \Sigma^R = - i \Sigma_{(a)}^R  - i \Sigma_{(b)}^R 
\end{equation}
where
\begin{eqnarray}
- i \Sigma_{(a)}^R &=& \frac{g^2}{4 a^6} | \Phi_0 |^2 \bigg[2 G^A (y', y) G^K (y, y') +
2 G^R (y, y') G^K (y', y) \nonumber\\
& & + 2 G^R (y, y') G^K (y, y')\bigg] \label{sigmaaR}\\
- i \Sigma_{(b)}^R &=& i \frac{g^2}{8 a^6} \bigg[ 2 G^R (y', y) G^A (y, y') G^R (y,
y') + 4 G^K (y', y) G^K (y, y') G^R (y, y') \nonumber\\
& & + 2 G^A (y', y) G^K (y, y') G^K
(y, y') + 2 G^A (y', y) G^A (y, y') G^A (y, y') \nonumber\\
& & + 2 G^A (y', y) G^R (y, y') G^R (y, y') \bigg] \label{sigmabR}
\end{eqnarray}
For each term, we take a Wigner transform, considering the slow variation of $a$ and $\Phi_0$ and follow similar steps to the computation of the condensate self-energy, considering the symmetrization of the last term in $\Sigma_{(a)}^R$ and in the first one for $\Sigma_{(b)}^R$ to get
\begin{eqnarray}
\Im (\Sigma_{(a)}^R) &=& \frac{\pi g^2}{2 a^6} \underset{\mathbf{p}}{\sum} | \Phi_0
|^2 \bigg[2 \delta (\varepsilon_{\mathbf{q}} + \varepsilon_{\mathbf{p}} -
\varepsilon_{\mathbf{k}} - \varepsilon_{\mathbf{p} + \mathbf{q} -
\mathbf{k}}) (F (p) - F (p + q - k)) \nonumber\\
& & - \delta (\varepsilon_{\mathbf{q}} +
\varepsilon_{\mathbf{k}} - \varepsilon_{\mathbf{p}} -
\varepsilon_{\mathbf{k} + \mathbf{q} - \mathbf{p}}) (F (p) + F (k + q - p)) \bigg]\label{sigmaaRfin}\\
\Im (\Sigma_{(b)}^R) &=& \frac{\pi g^2}{2 a^6} \underset{\mathbf{p},
\mathbf{q}}{\sum} \delta (\varepsilon_{\mathbf{p}} +
\varepsilon_{\mathbf{k} - \mathbf{q}} - \varepsilon_{\mathbf{k}} -
\varepsilon_{\mathbf{p} - \mathbf{q}}) \nonumber\\
& & \times \bigg[F (k - q) F (p) + 1 - F (p - q) (F
(k - q) + F (p))\bigg]\label{sigmabRfin}
\end{eqnarray}

\subsection{Particle self-energy $\Sigma^K$}

\begin{figure}[t]
    \centering   \includegraphics[width=1.2\linewidth,]{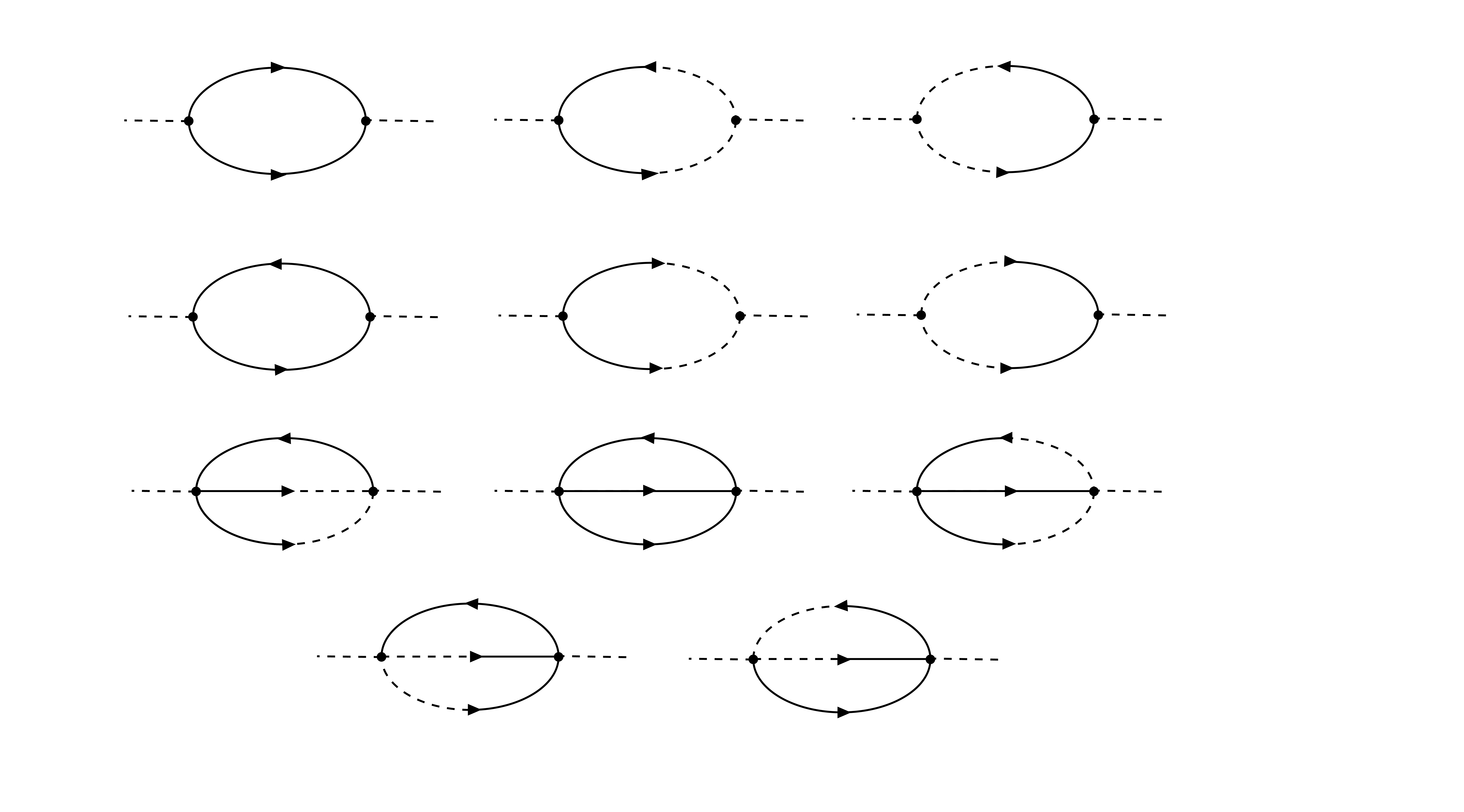}
    \caption{Feynman diagrams for the particle Keldysh self-energy at order $g^2$. The six upper diagrams correspond to $\Sigma^{K}_{(a)}$, the remaining corresponds to $\Sigma^{K}_{(b)}$.
    }
    \label{fig:sigmak}
\end{figure}

The contributions to the particle self-energy $\Sigma^K$ are depicted in figure \ref{fig:sigmak} and are given by
\begin{equation}
- i \Sigma^K =  - i \Sigma_{(a)}^K - i \Sigma_{(b)}^K
\end{equation}
where
\begin{eqnarray}
- i \Sigma_{(a)}^K &=& \frac{g^2}{4 a^6} | \Phi_0 |^2 \bigg[ 2 G^K (y, y')^2 + 4
G^A (y', y) G^R (y, y') + 4 G^R (y', y) G^A ( y, y' ) \nonumber\\
& & + 4 G^K (y, y') G^K (y', y) + 2 G^R (y, y')^2 + 2 G^A (y, y')^2 \bigg] \label{sigmaaK}\\
- i \Sigma_{(b)}^K &=& i \frac{g^2}{4 a^6} \bigg[ 2 G^K (y', y) G^R (y, y')^2 + 2
G^K (y', y) G^K (y, y')^2 \nonumber\\
& & + 4 G^A (y', y) G^K ( y, y' ) G^R (y, y')
+ 2 G^K (y', y) G^A (y, y')^2 \nonumber\\
& & + 4 G^R (y', y) G^A (y, y') G^K (y, y') \bigg] \label{sigmabK}
\end{eqnarray}

We use $G^R (y', y) G^A (y, y') + G^A (y', y) G^R (y, y') = - (G^R (y',
y) - G^A (y', y)) (G^R (y, y') - G^A (y, y'))$ and $G^A (y, y')^2 + G^R (y,
y')^2 = (G^R (y, y') - G^A (y, y'))^2$. Then, we apply a Wigner transform, and as before, symmetrizing the last term in $\Sigma_{(b)}^K$ we get
\begin{eqnarray}
\Sigma_{(a)}^K &=& \frac{\pi g^2}{2 i a^6} \underset{\mathbf{p}}{\sum} | \Phi_0
|^2 \bigg[4 \delta (\varepsilon_{\mathbf{q}} + \varepsilon_{\mathbf{p}} -
\varepsilon_{\mathbf{k}} - \varepsilon_{\mathbf{p} + \mathbf{q} -
\mathbf{k}}) (F (p + q - k) F (p) - 1) \nonumber\\
& & + 2 \delta(\varepsilon_{\mathbf{q}} + \varepsilon_{\mathbf{k}} -
\varepsilon_{\mathbf{p}} - \varepsilon_{\mathbf{k} + \mathbf{q} -
\mathbf{p}}) (F (k + q - p) F (p) + 1) \bigg]\label{sigmaaKfin}\\ 
\Sigma_{(b)}^K &=& \frac{\pi g^2}{i a^6} \underset{\mathbf{p},
\mathbf{q}}{\sum} \delta (\varepsilon_{\mathbf{p}} +
\varepsilon_{\mathbf{k} - \mathbf{q}} - \varepsilon_{\mathbf{k}} -
\varepsilon_{\mathbf{p} - \mathbf{q}}) \bigg[F (p - q) (F (k - q) F (p) + 1) \nonumber\\
& & - F (p) - F (k - q) \bigg]\label{sigmabKfin}
\end{eqnarray}

\subsection{Collisional terms}

With the computed self-energies we can construct the collisional terms for the kinetic equation of the particles. With the help of \eqref{collisional} we can generate two kind of collisional terms:
\begin{eqnarray}
I_{a}[f(k)] &=& i \Sigma_{(a)}^K+2(2f(k)+1)\Im(\Sigma_{(a)}^R)\\
I_{b}[f(k)] &=& i \Sigma_{(b)}^K+2(2f(k)+1)\Im(\Sigma_{(b)}^R)
\end{eqnarray}
We use explicitly \eqref{sigmaaRfin}, \eqref{sigmabRfin}, \eqref{sigmaaKfin} and \eqref{sigmabKfin}, after using \eqref{relationf} in the equations above. Using \eqref{condensatedens} and after some algebra we obtain that
\begin{eqnarray}
I_{a} &=& 8 \pi \frac{g^2}{a^6} n_c \mathbf{\underset{\mathbf{p},
\mathbf{l}, k}{\sum}} \delta (\varepsilon_{\mathbf{q}} +
\varepsilon_{\mathbf{p}} - \varepsilon_{\mathbf{k}} -
\varepsilon_{\mathbf{l}}) \delta (\mathbf{l} - \mathbf{p} - \mathbf{q}
+ \mathbf{k}) \bigg( \delta (\mathbf{p} - \mathbf{r}) - \delta (\mathbf{k} -
\mathbf{r}) \nonumber\\ 
& & - \delta (\mathbf{l} - \mathbf{r}) \bigg) 
\bigg[(1 + f_p) f_k f_l - f_p (1 + f_k) (1 + f_l)\bigg] \\
I_{b} &=& 8 \pi \frac{g^2}{a^6} \underset{\mathbf{p}, \mathbf{q}, \mathbf{l}}{\sum}
\delta (\varepsilon_{\mathbf{p}} + \varepsilon_{\mathbf{q}} -
\varepsilon_{\mathbf{k}} - \varepsilon_{\mathbf{l}}) \delta (\mathbf{l}
- \mathbf{p} - \mathbf{q} + \mathbf{k}) \nonumber\\
& & \times \bigg[ f_p f_q (f_k + 1) (f_l + 1) -
f_k f_l (f_p + 1) (f_q + 1) \bigg]
\end{eqnarray}
where we note that $F = 2 f + 1$.

Finally, we note that the collisional term $I_a$ is related with \eqref{sigmacondRfin} the condensate self energy. Indeed, defining $R$ as
\begin{eqnarray}
\label{erre}
R &=& - \Im (\Sigma_{(cond)}^{R}) = 2 \pi \frac{g^2}{a^6}
\underset{\mathbf{p}, \mathbf{k}, \mathbf{l}}{\sum} \delta
(\varepsilon_{\mathbf{q}} + \varepsilon_{\mathbf{p}} -
\varepsilon_{\mathbf{k}} - \varepsilon_{\mathbf{l}}) \delta (\mathbf{q}
- \mathbf{k} + \mathbf{p} - \mathbf{l}) \nonumber\\
& & \times \bigg[ f_p (1 + f_k) (1 + f_l) - (1 + f_p) f_k f_l \bigg]
\end{eqnarray}
we observe that it satisfies
\begin{equation}
R = \frac{1}{4 n_c} \underset{\mathbf{r}}{\sum} I_a
\end{equation}

\bibliographystyle{unsrt}
\bibliography{Refs_MixedDM}

\end{document}